\documentclass[dvipsnames,onefignum,onetabnum]{siamart171218}
\usepackage[top=1.4in,bottom=1.275in,left=1.4in,right=1.4in]{geometry}

\usepackage[T1]{fontenc}
\usepackage[english]{babel}\usepackage{csquotes}\usepackage[activate={true,nocompatibility},tracking=true]{microtype}\usepackage{hyphenat}\usepackage{siunitx}
\usepackage{placeins}
\usepackage{cite}\usepackage[begintext=``, endtext='']{quoting}
\SetBlockEnvironment{quoting}
\SetBlockThreshold{1}

\usepackage{amsfonts}
\usepackage{amssymb}
\usepackage{amscd}
\usepackage{mathtools}
\usepackage[mathscr]{eucal}
\usepackage{calligra}
\DeclareMathAlphabet{\mathcalligra}{T1}{calligra}{m}{n}
\DeclareFontShape{T1}{calligra}{m}{n}{<->s*[1.1]callig15}{}
\usepackage{bm}
\usepackage{autonum}\usepackage[nice]{nicefrac}

\usepackage{tikz}
\usepackage{pgf}
\usepackage{pgfplots}
\usepackage{pgfplotstable}
\usepackage{shellesc}
\pgfplotsset{compat=newest}
\usetikzlibrary{spy}
\usetikzlibrary{shapes}
\usetikzlibrary{backgrounds}
\usetikzlibrary{shadows}
\usetikzlibrary{matrix}
\usepgfplotslibrary{fillbetween}

\usepackage{comment}
\usepackage{graphicx}
\usepackage[colorinlistoftodos,prependcaption,textsize=small]{todonotes}
\usepackage{cleveref}

\crefname{equation}{}{}
\usepackage{csvsimple}
\usepackage{booktabs}
\usepackage{environ}
\usepackage{adjustbox}
\usepackage{relsize}
\usepackage[font=small,figurewithin=section]{caption}\usepackage[subrefformat=parens,labelformat=simple]{subcaption}
\usepackage{afterpage}

\usepackage{algorithm} 
\usepackage{algpseudocode}

\let\min\relax \DeclareMathOperator*\min{\vphantom{p}min}
\let\max\relax \DeclareMathOperator*\max{\vphantom{p}max}
\let\subset\relax \DeclareMathOperator{\subset}{\subseteq}
 
\let\tilde\widetilde
\let\hat\widehat

\newcommand\fnurl[2]{\href{#1}{#2}\footnote{\scalebox{.8}{\url{#1}}}}

\newcommand{\shiftexp}{\sigma}

\newcommand{\integral}[4]{\int_{#1}^{#2} #3 \,\mathrm{d}#4}

\newcommand{\R}{\mathbb{R}} \newcommand{\N}{\mathbb{N}}  \ifdef{\C}{\renewcommand{\C}{\mathbb{C}}}{\newcommand{\C}{\mathbb{C}}}  \newcommand{\zI}{\mathrm{i}}              \newcommand{\conj}[1]{\overline{#1}}
\newcommand{\T}{{\raisebox{.1ex}[0ex][0ex]{$\scriptscriptstyle\mathsf{T}$}}}

\newcommand{\bmn}{\bm{n}}

\newcommand{\bms}{\bm{s}}

\newcommand{\bmx}{\bm{x}}

\newcommand{\sff}{\mathsf{f}}

\newcommand{\sfr}{\mathsf{r}}

\newcommand{\sfu}{\mathsf{u}}

\newcommand{\sfA}{\mathsf{A}}
\newcommand{\sfB}{\mathsf{B}}

\newcommand{\sfF}{\mathsf{F}}

\newcommand{\sfI}{\mathsf{I}}

\newcommand{\sfK}{\mathsf{K}}

\newcommand{\sfM}{\mathsf{M}}

\newcommand{\sfP}{\mathsf{P}}

\newcommand{\sfS}{\mathsf{S}}

\newcommand{\mcD}{\mathcal{D}}

\newcommand{\mcO}{\mathcal{O}}

\newcommand{\bfx}{\mathbf{x}}

\newcolumntype{?}{!{\vrule width 1.2pt}}

\makeatletter
\newsavebox{\measure@tikzpicture}
\NewEnviron{scaletikzpicturetowidth}[1]{	\def\tikz@width{#1}	\def\tikzscale{1}\begin{lrbox}{\measure@tikzpicture}		\BODY
	\end{lrbox}	\pgfmathparse{#1/\wd\measure@tikzpicture}	\edef\tikzscale{\pgfmathresult}	\BODY
}
\makeatother

\definecolor{color1}{rgb}{0, 0.4470, 0.7410}
\definecolor{color2}{rgb}{0.8500, 0.3250, 0.0980}
\definecolor{color3}{rgb}{0.9290, 0.6940, 0.1250}
\definecolor{color4}{rgb}{0.7060, 0.3840, 0.7650}
\definecolor{color5}{rgb}{0.4660, 0.6740, 0.1880}
\definecolor{color6}{rgb}{0.3010, 0.7450, 0.9330}
\definecolor{color7}{rgb}{0.6350, 0.0780, 0.1840}
\definecolor{color8}{rgb}{0.7410, 0.3800, 0.1840}

\pgfplotsset{
  log x ticks with fixed point/.style={
      xticklabel={
        \pgfkeys{/pgf/fpu=true}
        \pgfmathparse{exp(\tick)}        \pgfmathprintnumber[fixed relative, precision=3]{\pgfmathresult}
        \pgfkeys{/pgf/fpu=false}
      }
  },
  log y ticks with fixed point/.style={
      yticklabel={
        \pgfkeys{/pgf/fpu=true}
        \pgfmathparse{exp(\tick)}        \pgfmathprintnumber[fixed relative, precision=3]{\pgfmathresult}
        \pgfkeys{/pgf/fpu=false}
      }
  }
}

\tikzset{
  ashadow/.style={opacity=.25, shadow xshift=0.07, shadow yshift=-0.07},
}

\definecolor{CustomGreen}{RGB}{65,169,50}

\graphicspath{{./figures/}}

\allowdisplaybreaks

\title{Semi matrix-free twogrid preconditioners for the Helmholtz equation with near optimal shifts}

\author{Daniel~Drzisga\thanks{Lehrstuhl f\"ur Numerische Mathematik, Fakult\"at f\"ur Mathematik (M2), Technische Universit\"at M\"unchen, Garching bei M\"unchen (\email{drzisga@ma.tum.de}, \email{koeppl@ma.tum.de}, \email{wohlmuth@ma.tum.de})}
\and Tobias~K{\"o}ppl\footnotemark[1]
\and Barbara~Wohlmuth\footnotemark[1]}

\begin{document}

\maketitle

\begin{abstract}
	Due to its significance in terms of wave phenomena a considerable effort has been put into the design of preconditioners for the Helmholtz equation. 
    One option to derive a preconditioner is to apply a multigrid method on a shifted operator. In such an approach, the wavenumber is shifted by some imaginary value. This step is motivated by the observation that the shifted problem can be more efficiently handled by iterative solvers when compared to the standard Helmholtz equation. However, up to now, it is not obvious what the best strategy for the choice of the shift parameter is. It is well known that a good shift parameter depends sensitively on the wavenumber and the discretization parameters such as the order and the mesh size.    
    Therefore, we study the choice of a near optimal complex shift such that an FGMRES solver converges with fewer iterations. Our goal is to provide a map which returns the near optimal shift for the preconditioner depending on the wavenumber and the mesh size. In order to compute this map, a data driven approach is considered: We first generate many samples, and in a second step, we perform a nonlinear regression on this data. With this representative map, the near optimal shift can be obtained by a simple evaluation. Our preconditioner is based on a twogrid V-cycle applied to the shifted problem, allowing us to implement a semi matrix-free method. The performance of our preconditioned FGMRES solver is illustrated by several benchmark problems with heterogeneous wavenumbers in two and three space dimensions.
\end{abstract}


\section{Introduction}
\label{sec:introduction}

The Helmholtz equation plays an important role in the mathematical modeling of wave phenomena like the propagation of sound and light. Because of its importance, the Helmholtz equation is of large interest for both, analytical and numerical research. In this work, we deal with the efficient solving of the linear systems of equations resulting from standard finite element discretizations of the Helmholtz equation.

According to literature \cite{cocquet2021closed,Gander2015Applying}, there are several reasons why the numerical solution of these equation systems is a challenging task. One reason is that the solutions of the Helmholtz equation are oscillating on a scale of $1/k$, where $k$ denotes the wavenumber. The wavenumber $k$ is proportional to the frequency of the simulated waves. As a consequence, a large number of mesh nodes is required to resolve high frequency waves. On closer examination, it turns out that the required number of mesh nodes is proportional to $k^2$. Moreover, low-order methods suffer from pollution effects, which implies that $\mcO(k^2)$ mesh nodes are not sufficient to bound the discretization error as the wavenumber $k$ increases. This implies that very large systems of equations have to be solved if large wavenumbers are considered. A further difficulty is that for large wavenumbers, these linear systems of equations can be distinctly indefinite such that classical iterative solvers perform poorly \cite{cools2013local}. For instance, a direct application of multigrid methods yields unsatisfactory results, since it can be shown that the smoothers as well as the coarse grid corrections cause growing error components \cite{brandt1986multigrid,elman2001multigrid}. 

These observations motivate the design of more effective iterative solvers. As a consequence, great effort has been made to achieve this goal. An overview of different solution methods that have been tested in this context can be found in \cite{Ernst2011Why,Gander2019class}. Due to the fact that classical iterative solvers like the Jacobi method and multigrid methods are not appropriate for a direct application to the Helmholtz problem \cite{Cocquet2017How,Stolk2014Multigrid}, Krylov subspace methods like GMRES \cite{Ramos2020Two} or BiCGSTab \cite{erlangga2004class} have attracted more attention. This is motivated by the fact that these Krylov subspace methods converge even in the case of indefinite matrices. However, their convergence can be very slow without a sophisticated preconditioner \cite{liu2016recursive,TAUS2020109706,saad1986gmres}.

It turns out that a simple modification of the original Helmholtz problem forms the basis to derive an efficient preconditioner for a Krylov subspace method. This is achieved, e.g.,~by adding a complex shift to the square of the wave number resulting in a new partial differential equation (PDE). This PDE is referred to as the \emph{shifted Laplacian problem} or the \emph{shifted Laplacian preconditioner}. In the remainder of this work, we will use the term \emph{shifted Laplacian problem}, since a preconditioner is only obtained when applying a numerical solver like a multigrid method to the shifted PDE.

A crucial issue in this context is to determine the optimal shift denoted by $\varepsilon$ \cite{Cocquet2017How,Gander2015Applying}. It can be shown that for $\varepsilon \in \mcO(k)$, a multigrid method applied to the shifted Laplacian problem is a good preconditioner. On the other hand using a shift $\varepsilon \in \mcO(k^2)$ \cite{Cocquet2017How,erlangga2006novel}, the standard multigrid method shows optimal convergence for the discrete counterpart of the shifted Laplacian problem. This means that there is a gap between the choice $\varepsilon \in \mcO(k)$ and $\varepsilon \in \mcO(k^2)$. From this one can conclude that a shift of $\varepsilon \in \mcO(k^\shiftexp),\;\shiftexp \in \left[1,2\right]$ should yield a solver for the unshifted Helmholtz problem, which can be regarded as a compromise between a fast multigrid convergence and a good preconditioner. Obviously, it is of great interest to determine an appropriate exponent $\shiftexp \in \left[1,2\right]$ for a Krylov subspace method, such that the method requires a minimal number of iterations to converge. However, there is no analytical formula how to choose this $\shiftexp$.

The goal of this paper is to design a Krylov subspace method using a near optimal shift. As the outer iterative solver for the standard Helmholtz equation, we choose the FGMRES method \cite{Saad1993Flexible}. As the preconditioner for the outer solver, we use the standard twogrid solver \cite{trottenberg2000multigrid} applied to the shifted problem. The twogrid solver uses the damped Jacobi method as a smoother. This allows for a semi-matrix free implementation of the iterative solver, meaning that most of the required matrix vector products can be realized without accessing a stored global sparse matrix. The Helmholtz equation as well as the shifted Laplacian problem are discretized using standard $Q_p,\;p \in \left\{1,2,3\right\}$, finite elements. In order to circumvent the lack of knowledge on how to choose an optimal shift, we use a data driven approach such as in \cite{Greenfeld2019Learning,Heinlein2019Machine,Luz2020Learning}. Thereby, for a given constant wavenumber $k$ and a mesh size $h$, the near optimal exponent $\shiftexp_{opt} \in \left[1,2\right]$ for the complex shift is determined by an optimization method. For this purpose, we use the golden-section method \cite{Kiefer1953Sequential}. 

This procedure is repeated for a large number of samples with respect to $k$ and $h$. Having for each sample the optimal exponent $\shiftexp_{opt}$ at hand, a map is constructed, such that a near optimal shift can be obtained by just evaluating the map for a given wavenumber and mesh size. By means of such a map, an optimal FGMRES solver can be constructed in an efficient way, particularly, if a highly heterogeneous distribution of the wavenumber has to be considered. In order to support our numerical findings, we perform a Local Fourier Analysis (LFA) for the $Q_1$-discretization in two space dimensions. As in \cite{cools2013local,erlangga2006novel}, we determine for the shift the convergence region of our twogrid solver. Thereby one has to note that in \cite{cools2013local,erlangga2006novel} finite difference stencils in two space dimensions are considered. Moreover, we restrict ourselves to the case $kh<0.75$ since our experimental results showed that such a mesh size can resolve the waves in a satisfactory manner and this kind of bound is backed by the theoretical analysis given in \cite{esterhazy2014analysis}. In particular, it is of interest to compute the exponents $\sigma_{MG} \in \left[1,2\right]$ marking the limit between convergence and divergence for the twogrid solver. Comparing $\sigma_{MG}$ and $\shiftexp_{opt}$, we can determine under which conditions the LFA analysis yields meaningful constraints for the choice of our samples. This can provide the basis for a more efficient training phase, since the different samples can be chosen a priori in the convergence domain of the twogrid method.

The rest of this paper is organized as follows: \Cref{sec:problem_setting} contains the problem setting as well as the variational formulations of the Helmholtz equation and the shifted Laplacian equation. Furthermore, the standard finite element discretizations and theoretical results from literature are recalled. \Cref{sec:LFA} focuses on the twogrid preconditioner. Using an LFA, we try to determine for which exponents $\shiftexp \in \left[1,2\right]$, the twogrid solver converges. Then in \Cref{sec:Multigrid}, the FGMRES solver combined with the twogrid solver is introduced. In \Cref{sec:data_generation}, the generation of training data that are used to estimate the optimal shift is described. The numerical results are presented and discussed in \Cref{sec:results}. Finally in \Cref{sec:Conclusion}, we summarize our main findings and give an outlook.
\section{Problem setting and variational formulations}
\label{sec:problem_setting}
The basic equation for modeling wave phenomena like the propagation of sound and light, is the linear wave equation
\begin{equation}
\label{eq:wave_equation}	
w_{tt} - c^2 \Delta w = \tilde{f},
\end{equation}
where the solution variable $w$ represents, e.g.,~the intensity of sound or light. The term $c$ denotes the propagation speed in a specific medium, and $\tilde{f}$ incorporates external source terms. Considering only time-harmonic solutions
$$
w(\bfx,t) = u(\bfx) e^{-\zI\omega t},
$$
with an angular frequency $\omega \in \mathbb{R}$, \eqref{eq:wave_equation} can be transformed into a stationary equation which is known as the Helmholtz equation \cite{ernst2013multigrid}:
\begin{equation}
\label{eq:Helmholtz}
\begin{alignedat}{3}
-\Delta u - k^2 u &= f
\qquad
&&\text{in } \Omega
,
\\
\frac{\partial u}{\partial \bmn} - \zI k u &= g
\qquad
&&\text{on } \partial \Omega
.
\end{alignedat}
\end{equation}
The parameter $k$ is given by $k=\omega/c$ and is referred to as the wavenumber. In the remainder of this work, we assume that it is given by a spatially varying function $k \colon \Omega \rightarrow \R$, where $\Omega \subset \R^d,\;d \in \left\{2,3\right\}$, represents a square or cubic domain. The source term $f$ results from the transformation of $\tilde{f}$. The Helmholtz equation is equipped with an impedance boundary condition, i.e.,~a first-order absorbing boundary condition with $g \colon \partial \Omega \rightarrow \C$. Using this notation, the \emph{shifted Laplacian problem} can be defined as follows:
\begin{equation}
\label{eq:Helmholtz_shift}
\begin{alignedat}{3}
-\Delta u - (k^2 + \zI \varepsilon) u &= f
\qquad
&&\text{in } \Omega
,
\\
\frac{\partial u}{\partial \bmn} - \zI k u &= g
\qquad
&&\text{on } \partial \Omega
.
\end{alignedat}
\end{equation}
Thereby the parameter $\varepsilon \in \R$ represents the imaginary shift with respect to the Helmholtz equation \eqref{eq:Helmholtz}.
For $\varepsilon \geq 0$, $f\in L^2( \Omega )$, $g \in L^2( \Gamma )$, and $k \in L^\infty(\Omega)$, the standard variational formulation of \eqref{eq:Helmholtz_shift} reads as follows: Find
$u \in H^1(\Omega,\mathbb{C})$ such that
\begin{equation}
\label{eq:var_form}
a_\varepsilon(u,v) = F(v),\; \text{ for all } v \in  H^1(\Omega,\mathbb{C}).
\end{equation}
The space $H^1(\Omega,\mathbb{C})$ consists of complex valued functions, whose real and imaginary parts are in the real valued space $H^1(\Omega)$. The sesquilinear form and the linear form of the variational formulation are given by:
\begin{align}
	a_\varepsilon(u,v) &= \integral{\Omega}{}{\nabla u \cdot \nabla \conj{v}}{\bmx}
	-\integral{\Omega}{}{(k^2 + \zI \varepsilon ) u \conj{v}}{\bmx} -\zI \integral{\partial \Omega}{}{k\, u \conj{v}}{\bmx}\\
	&= a(u,v) - m(u,v; k, \varepsilon) -\zI b(u,v;k)
\end{align}
and
\begin{align}
	F(v) &= \integral{\Omega}{}{f \conj{v}}{\bmx} + \integral{\partial \Omega}{}{g \conj{v}}{\bmx}.
\end{align}
Note that the sesquilinear form for the original Helmholtz problem is given by $a_0(\cdot,\cdot)$. According to \cite[Prop.~8.1.3]{Melenk1995generalized} the variational formulation \eqref{eq:var_form} is well-posed. 

In order to solve this equation numerically, standard $Q_p$-elements, $p \in \left\{1,2,3\right\}$, are used. Since we have assumed that $\Omega$ is a square or cubic domain, it can be decomposed into quadrilateral or hexahedral elements without approximation inaccuracies at the boundary $\partial\Omega$. The mesh size of the resulting grid is denoted by $h$. Further, let $V_h$ be the finite dimensional subspace of $H^1(\Omega,\mathbb{C})$ governed by $Q_p$-elements. Using this notation, the discrete version of \eqref{eq:var_form} has the following shape: Find $u_h \in V_h$ such that:
\begin{equation}
\label{eq:disc_var_form}
a(u_h,v_h) - m(u_h,v_h;k,\varepsilon) -\zI b(u_h,v_h;k) = F(v_h)
\text{ for all } v_h \in V_h.
\end{equation}
As it has been shown in \cite[Prop.~2.1]{esterhazy2014analysis}, the discrete variational Helmholtz formulation has a unique solution. Taking $u_h = \sum_{i=1}^N \sfu_i \phi_i$, where $\{\phi_i\}_{i=1}^N$ is a basis for $V_h$, problem \eqref{eq:disc_var_form} induces the following matrix equation for the coefficient vector $\sfu = [\sfu_1,\sfu_2,\ldots,\sfu_N]^\T$:
\begin{equation}
\label{eq:discreteproblem}
\sfK \sfu - \sfM(k,\varepsilon) \sfu - \zI \sfB(k) \sfu
=
{\sfF}
,
\end{equation}
where $\sfK_{ij} = a(\phi_j,\phi_i)$, $\sfM_{ij}(k,\varepsilon) = m(\phi_j,\phi_i;k,\varepsilon)$, $\sfB_{ij}(k) = b(\phi_j,\phi_i;k)$, and $\sfF_i = \integral{\Omega}{}{f \conj{\phi_i}}{\bmx} + \integral{\partial \Omega}{}{g \conj{\phi_i}}{\bmx}$.
We define the system matrix as
\begin{equation}
\label{eq:sys_matrix}
\sfA(k,\varepsilon) = \sfK - \sfM(k,\varepsilon) - \zI \sfB(k).
\end{equation}
\section{Local Fourier analysis for the twogrid solver in two space dimensions}
\label{sec:LFA}
In this section, we study the convergence behavior of the twogrid solver applied to the discrete equation \eqref{eq:disc_var_form}. Thereby only the two dimensional problem and the $Q_1$-discretization is considered.
Furthermore, we investigate only the local convergence of the twogrid solver i.e. we neglect the boundary conditions and consider only the following matrix:
\begin{equation}
\label{eq:local_operator}
\sfA(k,\varepsilon) = \sfK - (k^2 + \zI \varepsilon ) \sfM
\end{equation}
Of particular interest is the region of convergence of the twogrid solver. We follow the steps of a  Local Fourier Analysis (LFA) provided in \cite{cools2013local,trottenberg2000multigrid}. Therefore, we consider in the remainder of this work the system matrices $\sfA(k,k^\sigma),\;\sigma \in \left[1,2\right]$. The matrices $\sfA(k,k^\sigma)$ in \eqref{eq:local_operator} can be represented in a simplified way by means of the stencil notation:
$$
L_{h}( \sigma ) = 
\frac 13 
\left[ \begin{matrix} -1 & -1 & -1 \\ -1 & 8 & -1 \\ -1 & -1 & -1 \end{matrix} \right]
-( k^2 + \zI k^\sigma ) \frac{h^2}{36}
\left[ \begin{matrix} 1 & 4 & 1 \\ 4 & 16 & 4 \\ 1 & 4 & 1 \end{matrix} \right].
$$
The first part of the stencil $L_{h}( \sigma )$ corresponds to the standard stiffness matrix, while the second part represents the standard mass matrix. Let $\mathbf{G}_h$ be the grid given by
$$
\mathbf{G}_h = \left\{ \mathbf{x} = ( x_1,x_2 ) = ( k_1h,k_2h ),\; 
\mathbf{k} = ( k_1,k_2 ) \in \mathbb{Z}^2 \right\},
$$
and $V$ the index set of compact stencils 
$$
V = \left\{ \left. \kappa= (k_1,k_2) \right| k_1,\;k_2 \in \left\{-1,0,1\right\}\right\}\subset \mathbb{Z}^2.
$$
The values $s_\kappa \in \mathbb{C}$ are the coefficients of the stencil $S$. For simplicity, we have assumed that the $Q_1$-discretization is based on a mesh consisting of squares with an edge length of $h$.
The stencil $S$ applied to a grid function $w_h$ works as follows \cite[Chapter 4]{trottenberg2000multigrid}:
$$
S w_h( \mathbf{x} ) = \sum_{\kappa \in V} s_{\kappa} w_h( \mathbf{x} +h\kappa),\; \mathbf{x} \in \mathbf{G}_h.
$$

In a next step, we introduce the notation for the operator of the twogrid solver $T_h^{2h}$ \cite{briggs2000multigrid}. Applying $T_h^{2h}$ to the error function $e_h^l$ of the $l$-th iteration yields:
$$
e_h^{l+1} = T_h^{2h} e_h^l,\; T_h^{2h} = S_h^{\nu_2} K_h^{2h} S_h^{\nu_1},\;
K_h^{2h} = I_h - I_{2h}^h L_{2h}^{-1} I_h^{2h} L_{h}.
$$
Thereby, $S_h$ denotes the smoothing operator. In our work, we use the damped $\omega$-Jacobi smoother, where $\omega$ is the damping factor and $\nu_j,\;j\in \left\{1,2 \right\}$, are the number of pre- and postsmoothing steps. 
It is a well known fact that the operator for the damped Jacobi is given by
$$
S_h( \sigma ) = I_h - \omega D_h^{-1}( \sigma ) L_h( \sigma ), 
$$
where $I_h$ is the identity operator, and $D_h( \sigma )$ corresponds to the diagonal of the matrix in \eqref{eq:local_operator}. A straightforward computation yields the following stencil for $S_h( \sigma )$:
$$
S_h( \sigma ) = \left[ \begin{matrix} \omega \frac{1 + \lambda h^2 \frac{1}{3}}{8 - \lambda h^2 \frac{4}{3}} & \omega \frac{1 + \lambda h^2 \frac{1}{3}}{8 - \lambda h^2 \frac{4}{3}} & \omega \frac{1 + \lambda h^2 \frac{1}{3}}{8 - \lambda h^2 \frac{4}{3}} \\ 
\omega \frac{1 + \lambda h^2 \frac{1}{3}}{8 - \lambda h^2 \frac{4}{3}} & 1-\omega & \omega \frac{1 + \lambda h^2 \frac{1}{3}}{8 - \lambda h^2 \frac{4}{3}} \\ 
\omega\frac{1 + \lambda h^2 \frac{1}{3}}{8 - \lambda h^2 \frac{4}{3}} & \omega \frac{1 + \lambda h^2 \frac{1}{3}}{8 - \lambda h^2 \frac{4}{3}} & \omega \frac{1 + \lambda h^2 \frac{1}{3}}{8 - \lambda h^2 \frac{4}{3}}
\end{matrix} \right],
$$
where we have used the abbreviation $\lambda = k^2 + \zI k^\sigma$. 
Combining the smoothing operator with the correction operator $K_h^{2h}$, results in the twogrid operator $T_h^{2h}$. $K_h^{2h}$ itself is given by a combination of $L_{2h}$, $I_h^{2h}$, and $I_{2h}^h$.
The operators $$
I_h^{2h}: \mathbf{G}_h \rightarrow \mathbf{G}_{2h} \text{ and } I_{2h}^h: \mathbf{G}_{2h} \rightarrow \mathbf{G}_{h}
$$ 
stand for the restriction and prolongation where $\mathbf{G}_{2h}$ denotes the coarse grid.

According to \cite[Chapters 2 and 4]{trottenberg2000multigrid}, the stencils for the prolongation, restriction and identity operator are given by:
$$
I_{2h}^h = \frac 14 \left] \begin{matrix} 1 & 2 & 1 \\ 2 & 4 & 2 \\ 1 & 2 & 1
 \end{matrix} \right[,\;
I_h^{2h} = \frac 14 \left[ \begin{matrix} 1 & 2 & 1 \\ 2 & 4 & 2 \\ 1 & 2 & 1
\end{matrix} \right] \text{, and } 
I_h  = \left[ \begin{matrix} 0 & 0 & 0 \\ 0 & 1 & 0 \\ 0 & 0 & 0
\end{matrix} \right].
$$
$L_{2h}$ in stencil notation has a similar shape as $L_{h}$, the only difference is that $h$ has to be replaced by $2h$, and $\mathbf{x}$ is an element of the coarse grid $\mathbf{G}_{2h}$. The inverted brackets for the stencil of the prolongation operator $I_{2h}^h$ indicate that this stencil has to be applied in a different way as the remaining stencils, see \cite[Chapter 2]{trottenberg2000multigrid} for more details on the notation. Applying $I_{2h}^h$ to a coarse grid function $w_{2h}$ yields a fine grid function $w_{h}$ using the following rule \cite[Chapter 2]{trottenberg2000multigrid}: 
$$
w_{h}( \mathbf{x} +h\kappa) = (I_{2h}^h )_{\kappa} w_{2h}( \mathbf{x} ) ,\;\mathbf{x} \in \mathbf{G}_{2h},
$$
where $(I_{2h}^h )_{\kappa}$ is the entry of the stencil $I_{2h}^h$ belonging to the index $\kappa \in V$.

The convergence analysis by means of LFA is based on the assumption that the error function $e_h^l$ on the fine grid $\mathbf{G}_h$ can be represented by a linear combination of Fourier modes \cite{trottenberg2000multigrid}:
$$
\varphi_h( \mathbf{\theta},\mathbf{x} ) = e^{\zI \mathbf{\theta} \cdot \mathbf{x}/h}
=  e^{\zI \theta_1 x_1/h} e^{\zI \theta_2 x_2/h},\;\mathbf{x} \in \mathbf{G}_h.
$$
$\theta = ( \theta_1,\theta_2 ) \in \mathbb{R}^2$ denote the Fourier frequencies, which may be restricted to the domain $\Theta = \left(-\pi,\pi\right]^2 \subset \mathbb{R}^2$, see \cite{cools2013local}, as a consequence of the fact that 
$$
\varphi_h( \mathbf{\theta} + 2\pi,\mathbf{x} ) = \varphi_h( \mathbf{\theta},\mathbf{x} ).
$$
Thus the space of Fourier modes is given by:
$$
E_h = \text{span}\left\{  \left. \varphi_h( \mathbf{\theta},\mathbf{x} ) = e^{\zI \mathbf{\theta} \cdot \mathbf{x}/h} \right| \mathbf{x} \in \mathbf{G}_h,\; \mathbf{\theta} \in \Theta \right\}.$$
According to \cite{trottenberg2000multigrid}, the Fourier modes are the formal eigenfunctions of an operator 
$$
H \in \left\{ T_h^{2h},\;I_h,\;I_h^{2h},\;I_{2h}^h,\;L_{h},\;L_{2h},\;S_h \right\}.
$$
This means that:
$$
H \varphi_h( \mathbf{\theta},\mathbf{x} ) = \tilde{H}( \mathbf{\theta} ) \varphi_h( \mathbf{\theta},\mathbf{x} ), 
$$
where $\tilde{H}( \mathbf{\theta} )$ is called the Fourier symbol of the operator $H$. In a next step, we decompose the frequency space into a low frequency space $T^{low} = \left(-\frac{\pi}{2},\frac{\pi}{2}\right]^2$ and 
a high frequency space $T^{high} = \Theta \setminus T^{low}$. For each low frequency $\mathbf{\theta} \in T^{low} = \left( -\frac{\pi}{2}, \frac{\pi}{2} \right]^2,$ we define the following four frequencies:
$$
\mathbf{\theta}^{(0,0)} = ( \theta_1,\theta_2 ),\;
\mathbf{\theta}^{(1,1)} = ( \overline{\theta}_1,\overline{\theta}_2 ),\;
\mathbf{\theta}^{(1,0)} = ( \overline{\theta}_1,\theta_2 ),\;
\mathbf{\theta}^{(0,1)} = ( \theta_1,\overline{\theta}_2 ),
$$
where
$$
\overline{\mathbf{\theta}}_i =
\begin{cases}
\mathbf{\theta}_i + \pi \text{ if } \theta_i<0,\\
\mathbf{\theta}_i - \pi \text{ if } \theta_i \geq 0.
\end{cases}
$$
Obviously, the four frequencies fulfil
$$
\varphi_h( \mathbf{\theta}^\alpha,\mathbf{x} ) = \varphi_{2h}( 2\mathbf{\theta}^{(0,0)},\mathbf{x} ),\; \mathbf{x} \in \mathbf{G}_{2h},
\;\mathbf{\alpha} \in I = \left\{ (0,0),\;(1,1),\;(1,0),\;(0,1)\right\}.
$$ Fourier modes with this relationship are called \emph{harmonic to each other}. Considering a given low frequency $\mathbf{\theta} = \mathbf{\theta}^{(0,0)}$, we define a four dimensional space of harmonics by
$$
E_h^{\mathbf{\theta}} =
\text{span}\left\{ \left. \varphi( \mathbf{\theta}^{\alpha}, \cdot ) \right| 
\mathbf{\alpha} \in  I \right\}.
$$
An important feature of these spaces is that they are invariant under the twogrid operator $T_h^{2h}$. Applying $T_h^{2h}$ to an arbitrary element $\Psi \in E_h^{\mathbf{\theta}}$ which is represented by some coefficients $A^{\mathbf{\alpha}}$:
$$
\Psi = \sum_{\mathbf{\alpha} \in I} A^{\mathbf{\alpha}} \varphi( \mathbf{\theta}^{\alpha}, \cdot ),
$$
yields new coefficients $B^{\mathbf{\alpha}}$:
$$
\begin{pmatrix}
B^{(0,0)} \\ B^{(1,1)} \\ B^{(1,0)}  \\ B^{(0,1)} 
\end{pmatrix} 
= \hat{T}_h^{2h} ( \mathbf{\theta},\sigma )
\begin{pmatrix}
A^{(0,0)} \\ A^{(1,1)} \\ A^{(1,0)}  \\ A^{(0,1)} 
\end{pmatrix}. 
$$
$\hat{T}_h^{2h} ( \mathbf{\theta},\sigma ) \in \mathbb{C}^{4 \times 4}$ is a matrix depending on the frequency $\mathbf{\theta}$ and the exponent $\sigma$. It represents the operator $T_h^{2h}$ with respect to $E_h^{\mathbf{\theta}}$. In order to determine the convergence behavior of the twogrid method, we study how the coefficients $A^{\mathbf{\alpha}}$ are transformed by $\hat{T}_h^{2h} ( \mathbf{\theta},\sigma )$ into new coordinates $B^{\mathbf{\alpha}}$. Therefore, the local convergence factor $\rho_{loc}( \sigma )$  depending on the complex shift exponent in the twogrid operator is taken into account:
\begin{equation}
\label{eq:def_conv}
\rho_{loc}( \sigma  ) = \sup\left\{ \left. \rho( \hat{T}_h^{2h}( \mathbf{\theta},\sigma  ) ) \right| \mathbf{\theta} \in T^{low}, \mathbf{\theta} \notin \Lambda \right\}.
\end{equation}
Thereby $\Lambda \subset \left(-\pi,\pi\right]^2$ denotes the set of frequencies for which $\hat{L}_{2h}$ and $\hat{L}_{h}$ are not invertible and $\rho( \hat{T}_h^{2h}( \mathbf{\theta},\sigma ) )$ is the spectral radius of $\hat{T}_h^{2h}( \mathbf{\theta},\sigma )$. This means that finding the convergence factor for a fixed frequency is reduced to finding the spectral radius of a $4\times 4$ matrix. To obtain the convergence rate of the twogrid solver, one has to search for the maximum of this quantity in the low frequency space $T^{low}$. The search for the maximal spectral radius is restricted to $T^{low}$, since the pattern of $\rho( \hat{T}_h^{2h} )$ in $T^{low}$ is extended periodically to the whole frequency domain $\Theta$. Furthermore, an important feature of a multigrid solver is the damping of the low frequency error components \cite{trottenberg2000multigrid}.   

To compute $\hat{T}_h^{2h} ( \mathbf{\theta},\sigma )$, the operators occurring in the definition of $T_h^{2h}$ are represented with respect to the harmonic space $E_h^{\mathbf{\theta}}$ by means of $4\times 4$ matrices. Exceptions are the prolongation, restriction and $L^{-1}_{2h}$ operator whose matrices are given by $1\times 4$, $4\times 1$ and $1\times 1$ matrices, since these are mappings related to $E_h^{\mathbf{\theta}}$ and $E_{2h}^{\mathbf{\theta}}$. The latter space is the space of harmonics with respect to the coarse grid $\mathbf{G}_{2h}$:
$$
E_{2h}^{\mathbf{\theta}} =
\text{span}\left\{ \left. \varphi_{2h}( \mathbf{\theta}^{\alpha}, \cdot ) \right| 
\alpha \in  \left\{(0,0) \right\} \right\}.
$$
The matrices defined with respect to the harmonic spaces are indicated by a hat:
\begin{align}	
\hat{T}_h^{2h}( \mathbf{\theta},\sigma  )  &= \hat{S}_h^{\nu_2}( \mathbf{\theta},\sigma  ) \hat{K}_h^{2h}( \mathbf{\theta},\sigma ) \hat{S}_h^{\nu_1}( \mathbf{\theta},\sigma  ), \\
\ \\
\hat{K}_h^{2h}( \mathbf{\theta},\sigma )  &= \hat{I}_h - \hat{I}_{2h}^h( \mathbf{\theta} )  \hat{L}_{2h}^{-1}( 2\mathbf{\theta},\sigma  )  \hat{I}_h^{2h}( \mathbf{\theta} )  \hat{L}_{h}( \mathbf{\theta},\sigma  ).
\end{align}

\begin{figure}[H]
	\includegraphics[width=0.485\textwidth]{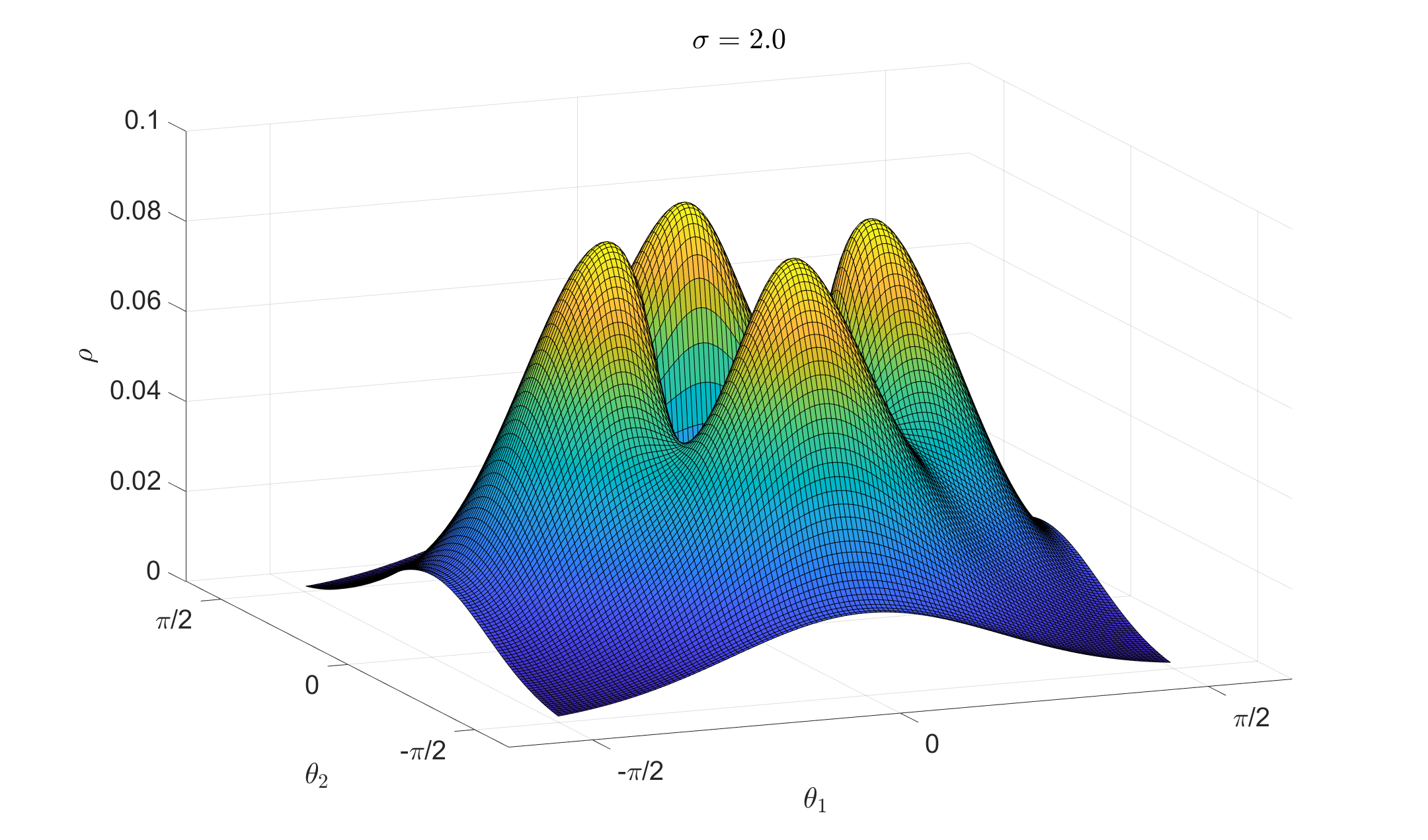}
    \includegraphics[width=0.485\textwidth]{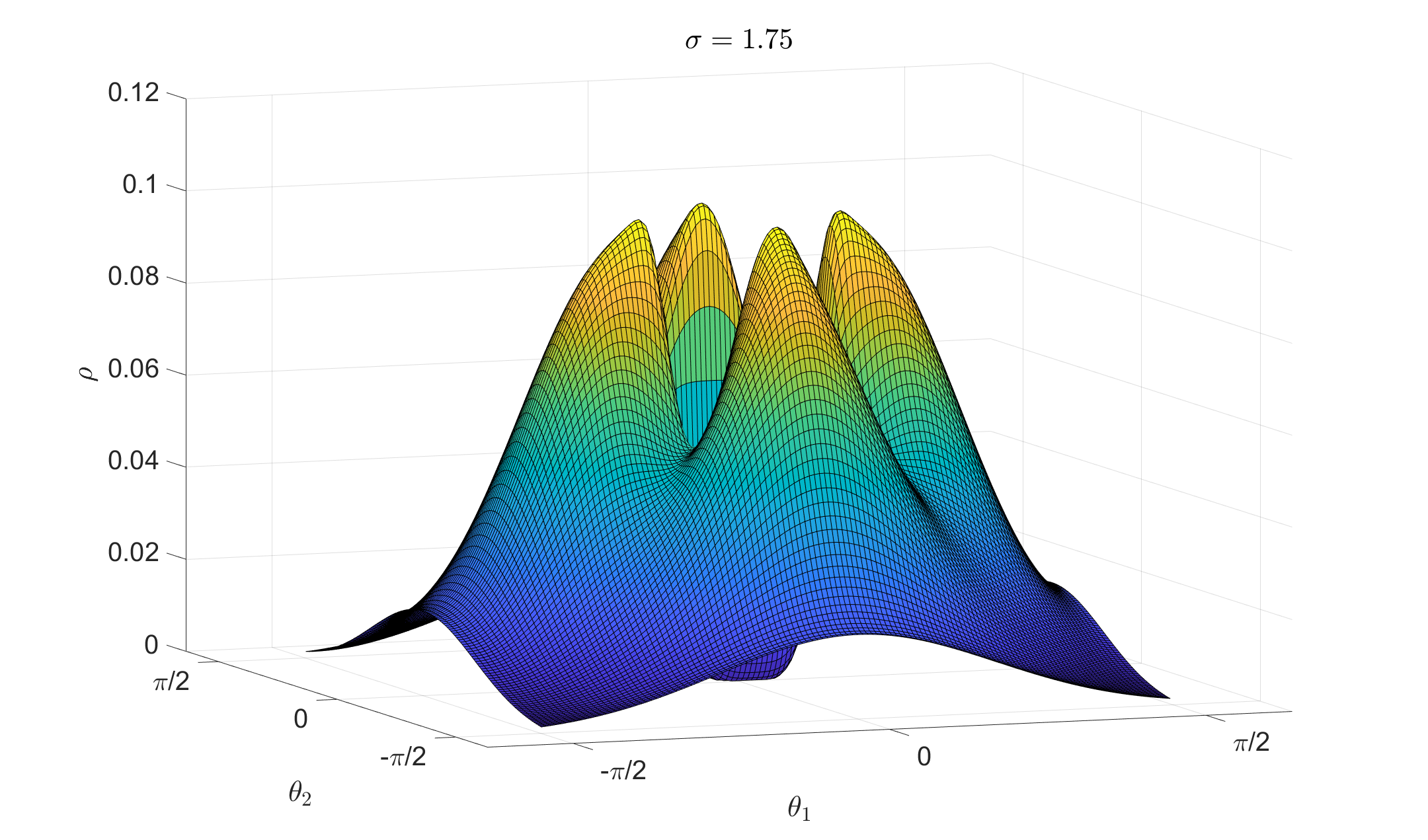}
    \includegraphics[width=0.485\textwidth]{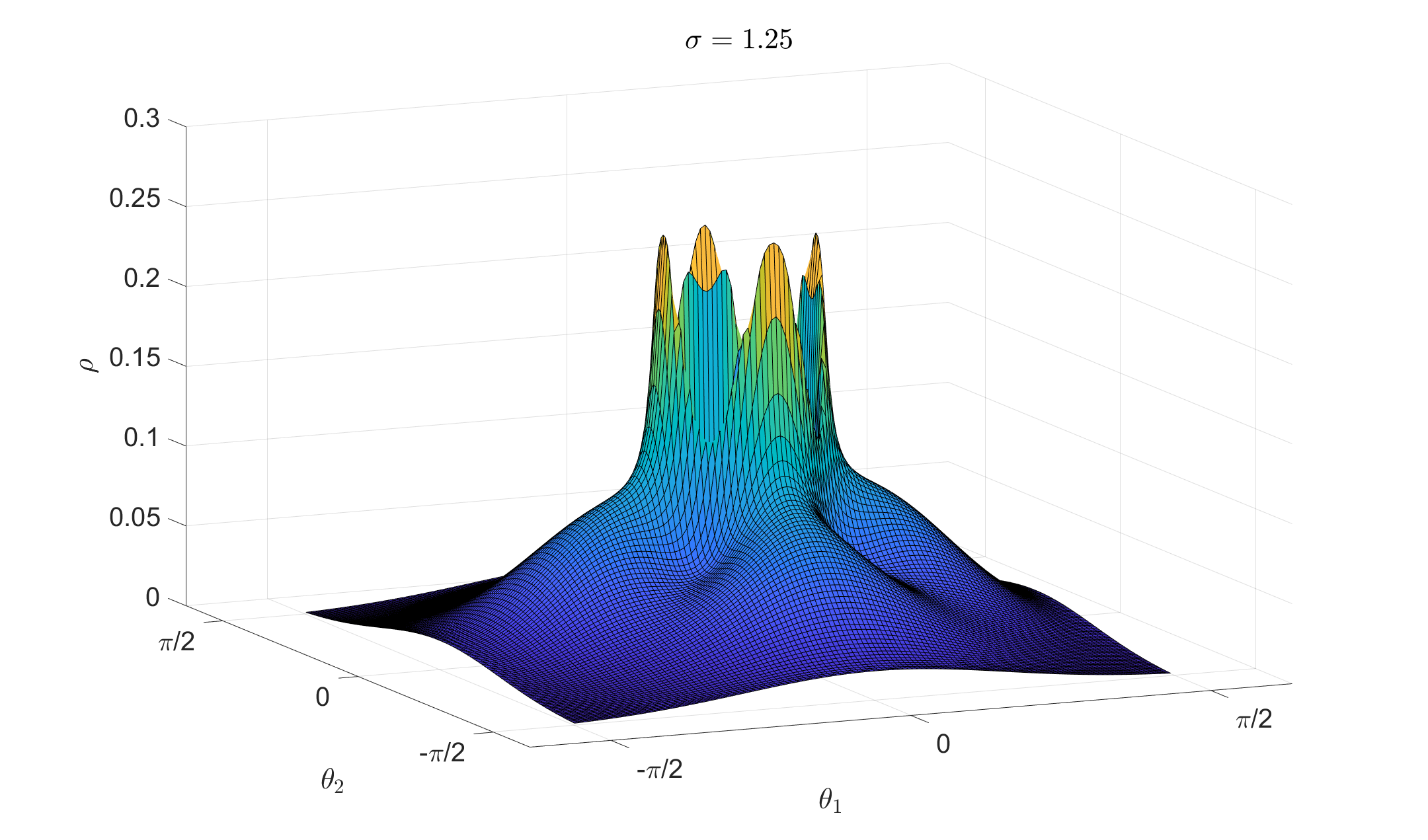}
    \includegraphics[width=0.485\textwidth]{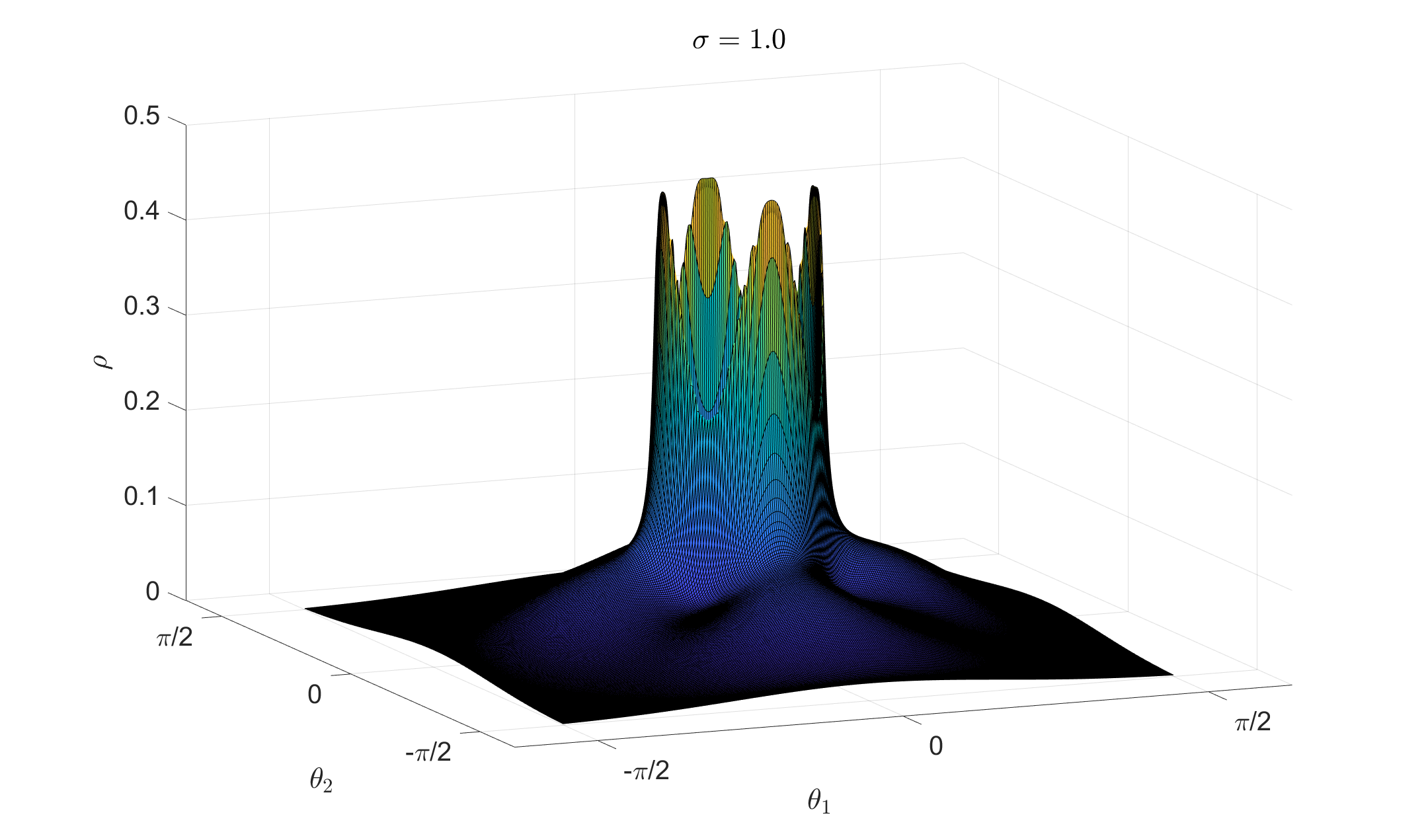}
\caption{\label{fig:Sigmas} Convergence factors $\rho$ in the frequency space $T^{low}$ in case of four different shifts $\sigma \in \left\{ 1.0,1.25,1.75,2.0 \right\}$.}	
\end{figure}

\begin{figure}[H]
	\includegraphics[width=0.485\textwidth]{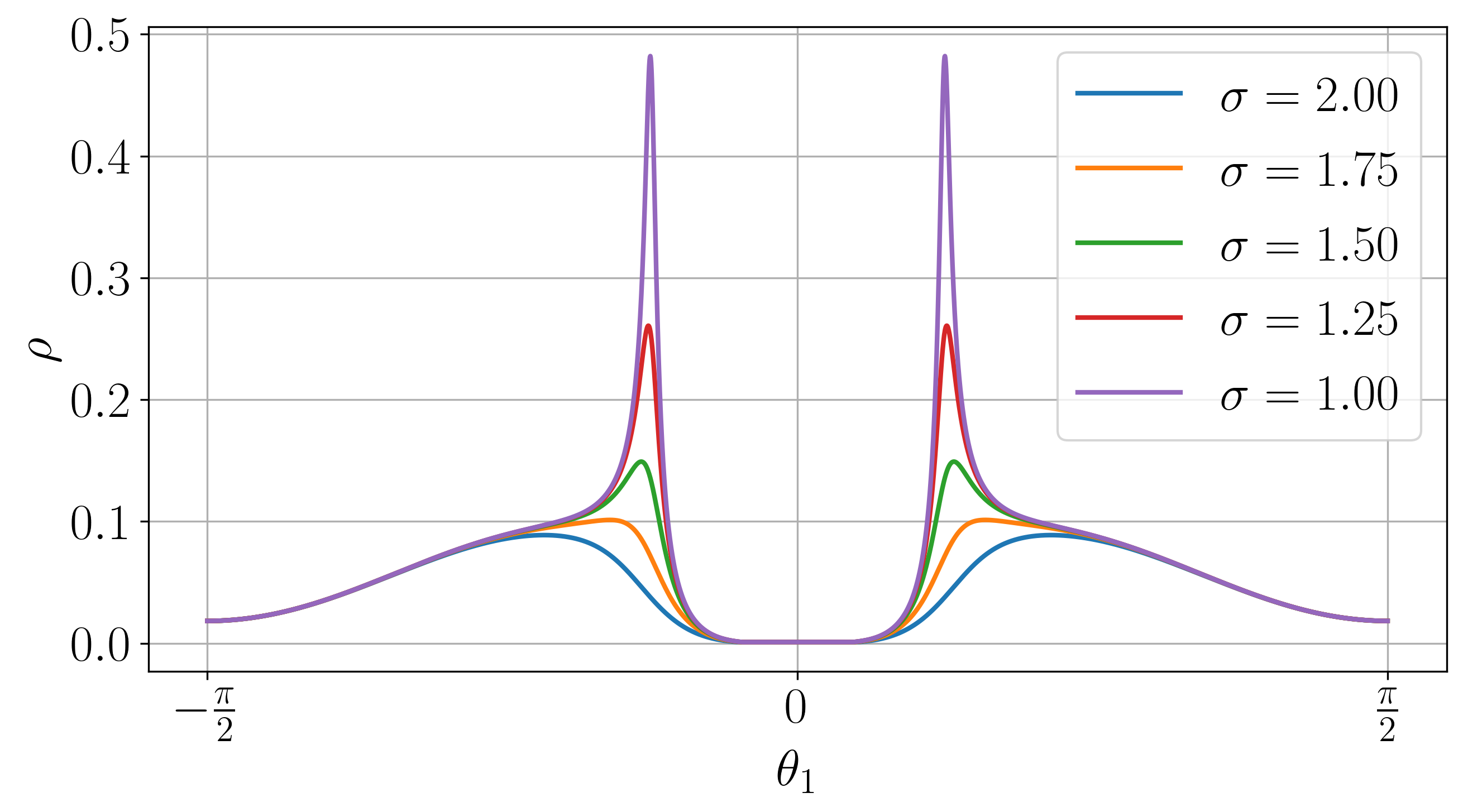}
	\includegraphics[width=0.485\textwidth]{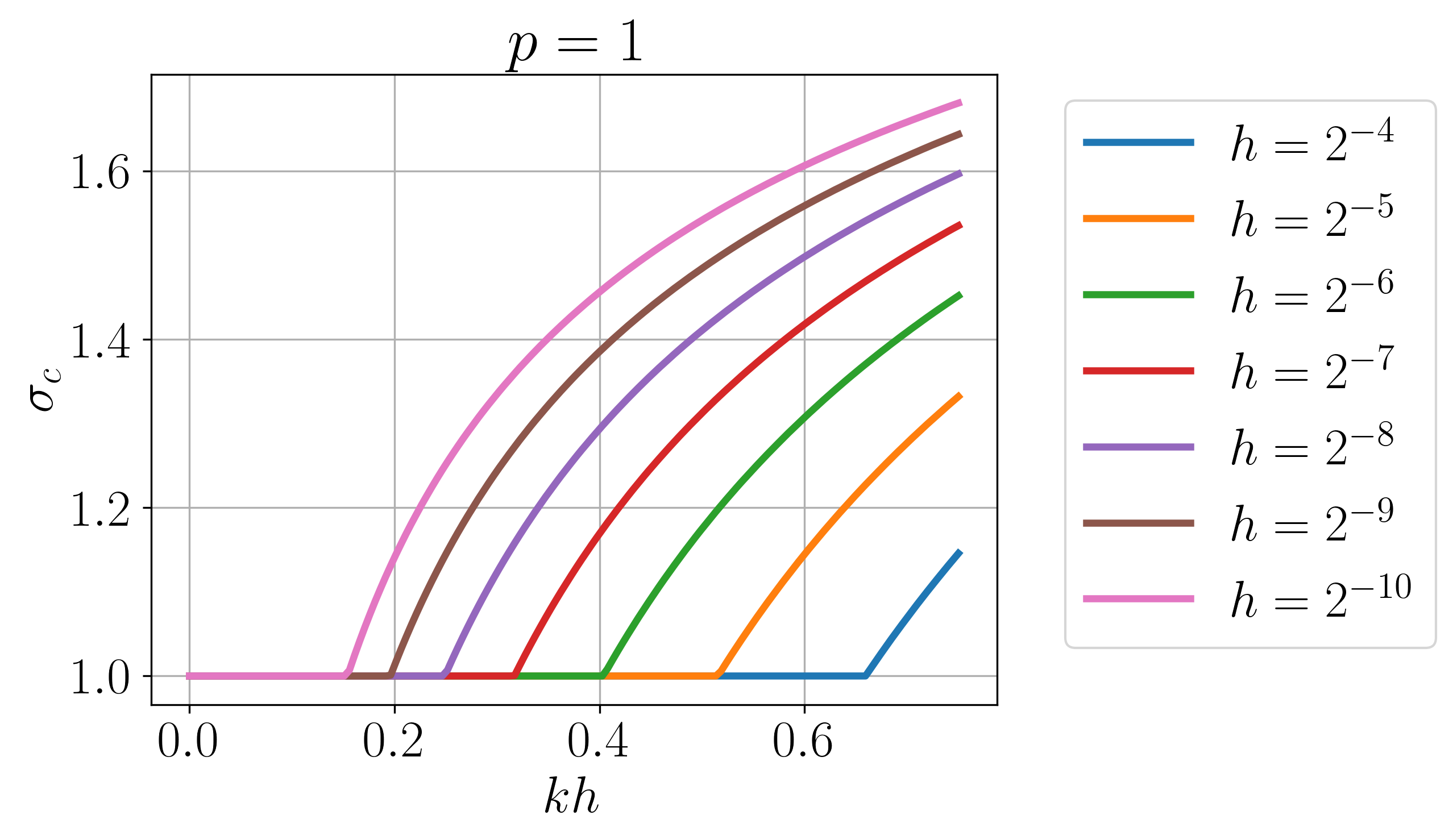}
	\caption{\label{fig:sigma_LFA} Left: Convergence factor $\rho$ for $\theta_2=0$ and $\theta_1 \in \left[-\frac{\pi}{2},\frac{\pi}{2} \right]$.The exponents for the shift are chosen as follows: $\sigma \in \left\{ 1.00,1.25,1.50,1.75,2.00 \right\}$, right: Minimal exponent $\sigma_c$ with respect to $kh$ for different mesh sizes $h \in \left\{2^{-4},\ldots,2^{-10}\right\}$.}
\end{figure}

In the following, we list the matrices with respect to the harmonic spaces. Obviously, the matrix for the identity operator $\hat{I}_h$ is given by the identity matrix. The discretization operator $L_h$ can also be represented by a diagonal matrix  \cite{cools2013local}\cite[Chapter 2]{trottenberg2000multigrid}:
$$
\hat{L}_h( \mathbf{\theta},\sigma ) = \text{diag}
\begin{pmatrix}
\tilde{L}_h( \mathbf{\theta}^{(0,0)}, \sigma ), & 
\tilde{L}_h( \mathbf{\theta}^{(1,1)}, \sigma ), &
\tilde{L}_h( \mathbf{\theta}^{(1,0)}, \sigma ), &
\tilde{L}_h( \mathbf{\theta}^{(0,1)}, \sigma )
\end{pmatrix}. 
$$
The symbol of $L_h$ is given by: 
\begin{align}
	\tilde{L}_h( \mathbf{\theta}, \sigma ) &= \frac 13 ( 8- 2 \cos (\theta_1) - 2 \cos (\theta_2) - 4 \cos (\theta_1) \cos (\theta_2) ) \\
	&-\frac{\lambda h^2}{36} (16+8\cos(\theta_1) +8\cos(\theta_2) + 4 \cos (\theta_1) \cos (\theta_2) ). 
\end{align}
The matrix for the coarse grid operator and $\mathbf{\theta} = \mathbf{\theta}^{(0,0)}$ is given by:
$$
\hat{L}_{2h}(\mathbf{\theta},\sigma) = \tilde{L}_{2h}(\mathbf{\theta},\sigma ),\;
L_{2h}(\sigma )\varphi_{2h}( 2\mathbf{\theta},\mathbf{x} ) =
\tilde{L}_{2h}(\mathbf{\theta},\sigma) \varphi_{2h}( 2\mathbf{\theta},\mathbf{x} ).
$$
The exact formula for $\tilde{L}_{2h}(\mathbf{\theta},\sigma )$ reads as follows:
\begin{align}
	\tilde{L}_{2h}(\mathbf{\theta},\sigma ) &= \frac 13 ( 8- 2 \cos (2\theta_1) - 2 \cos (2\theta_2) - 4 \cos (2\theta_1) \cos (2\theta_2) ) \\
	&-\frac{\lambda h^2}{9} (16+8\cos(2\theta_1) +8\cos(2\theta_2) + 4 \cos (2\theta_1) \cos (2\theta_2) ).
\end{align}
In a next step the matrices for the restriction operator and the prolongation operator are derived:
$$
\hat{I}_{2h}^h (\mathbf{\theta} ) = 
\begin{pmatrix}
\tilde{I}_{2h}^h( \mathbf{\theta}^{(0,0)} ) \\
\tilde{I}_{2h}^h( \mathbf{\theta}^{(1,1)} ) \\
\tilde{I}_{2h}^h( \mathbf{\theta}^{(1,0)} ) \\
\tilde{I}_{2h}^h( \mathbf{\theta}^{(0,1)} )
\end{pmatrix} =
\begin{pmatrix}
(1+\cos(\theta_1) ) (1+\cos(\theta_2) ) \\
(1-\cos(\theta_1) ) (1-\cos(\theta_2) ) \\
(1-\cos(\theta_1) ) (1+\cos(\theta_2) ) \\
(1+\cos(\theta_1) ) (1-\cos(\theta_2) )
\end{pmatrix} 
\text{ and } 
\hat{I}_h^{2h}(\mathbf{\theta} ) =  \frac 14 \cdot (\hat{I}_{2h}^h (\mathbf{\theta} ) )^T.
$$
Thereby the matrix entries are determined by the following equation:
$$
I_{2h}^h \varphi_{2h}( 2\mathbf{\theta},\mathbf{x} ) =
\sum_{\alpha} \tilde{I}_{2h}^h( \mathbf{\theta}^\alpha ) \varphi_{h}( \mathbf{\theta}^\alpha,\mathbf{x} ) \text{ and } 
I_h^{2h} \varphi_{h}( \mathbf{\theta}^\alpha,\mathbf{x} ) = \tilde{I}_h^{2h}(\mathbf{\theta}^\alpha) \varphi_{2h}( 2\mathbf{\theta}^{(0,0)},\mathbf{x} ),\; \alpha \in I.
$$
We point out that the restriction and prolongation operator are independent of $k$ and $\sigma$.
Summarizing the previous considerations, one obtains the matrix $\hat{K}_h^{2h}(\mathbf{\theta},\sigma )$. Finally, it remains to specify the matrix for the smoother. According to \cite[Chapter 4]{trottenberg2000multigrid}, it is given by a diagonal matrix:
$$
\hat{S}_{h}( \mathbf{\theta},\sigma ) = \text{diag}
\begin{pmatrix}
\tilde{S}_{h}( \mathbf{\theta}^{(0,0)},\sigma ), & 
\tilde{S}_{h}( \mathbf{\theta}^{(1,1)},\sigma ), &
\tilde{S}_{h}( \mathbf{\theta}^{(1,0)},\sigma ), &
\tilde{S}_{h}( \mathbf{\theta}^{(0,1)},\sigma )
\end{pmatrix}.
$$
A straightforward calculation yields for $\mathbf{\theta} = \mathbf{\theta}^{(0,0)}$:
\begin{align}
	\tilde{S}_{h}( \mathbf{\theta}, \sigma) = (1-\omega) 
	&+ \omega \frac{\frac 23 + \lambda h^2 \frac{8}{36}}{\frac 83 - \lambda h^2 \frac{16}{36}}
	\cos( \theta_1 ) \\
	&+ \omega \frac{\frac 23 + \lambda h^2 \frac{8}{36}}{\frac 83 - \lambda h^2 \frac{16}{36}}
	\cos( \theta_2 ) 
	+ \omega \frac{\frac 43 + \lambda h^2 \frac{4}{36}}{\frac 83 - \lambda h^2 \frac{16}{36}}
	\cos( \theta_1 ) \cos( \theta_2 ).
\end{align}
Finally, we have all ingredients at hand to assemble the matrix $\hat{T}_h^{2h}$ so that $\rho_{loc}( \sigma )$ can be computed  numerically. In \cref{fig:Sigmas}, we show some typical surfaces for $\rho$ in the frequency space $T^{low}$ for $\nu_1=\nu_2=3$, $\omega = 2/3$, $h=2^{-5}$ and $k=0.25 \cdot h$. It can be observed that the maximum of the spectral radius $\rho( \hat{T}_h^{2h} )$, i.e., $\rho_{loc}$ increases as $\sigma$ is decreased from $\sigma = 2$ towards $\sigma = 1$. In addition to that, there is a symmetry with respect to the axes of the coordinate system. This can be explained by the fact that the entries of the above matrices are composed of cosine functions. We further observe numerically that the maxima are located on the axes for $\theta_1 = 0$ or $\theta_2 =0$ (see left of \cref{fig:sigma_LFA}). Motivated by these observations, we restrict our search for the maximal spectral radius of $\hat{T}_h^{2h}$ to the line with $\theta_2 =0$.
Once calculated, the local convergence factor $\rho_{loc}( \sigma )$ helps us to determine the minimal exponent $\sigma_c$ for the shift $\varepsilon$ separating the interval $I_{\sigma} = \left[1,2\right]$ into two subsets $I_{conv}$ and $I_{div} = I_{\sigma} \setminus I_{conv}$, where 
$I_{conv} = \left[ 2,\sigma_c \right] \text{ and } \sigma_c = \text{argmin}_{\sigma \in I_{\sigma}} \left\{ \rho_{loc}( \sigma ) < 1 \right\}.$ 

In other words: $I_{conv} \subset I_{\sigma}$ contains the exponents $\sigma$ for which the twogrid method converges. Varying the smoothing steps $\nu_1$ and $\nu_2$ as well as the damping factor $\omega$ for $kh<0.75$ and a fixed mesh size $h$ shows only little impact on the graph of $\sigma_c$. On the other hand, varying the mesh size $h$ and keeping the remaining parameters fixed, shows a significant impact on $\sigma_c$ (see right of \cref{fig:sigma_LFA}). On closer examination, it can be observed that for a fixed wavenumber $k$ a small mesh size $h$ enlarges the interval $I_{conv}$. Reduced intervals $I_{conv}$ arise for a fixed $h$ and a sufficiently large $k$.

\section{Semi matrix-free FGMRES with twogrid shifted Laplacian preconditioner}
\label{sec:Multigrid}
In this section, we draw our attention to the twogrid method from the previous section being used as a preconditioner within a Krylov subspace method.
Therefore, we briefly describe the building blocks of the iterative solver used for the numerical experiments in the subsequent sections.
For the outer iterations of our solver, we use the preconditioned FGMRES Krylov subspace method \cite{Saad1993Flexible} to solve the system
$$\sfA_0 \sfP_\varepsilon^{-1} (\sfP_\varepsilon \sfu) = \sfF$$
for a matrix $\sfA_0$, a preconditioner $\sfP_\varepsilon$, and a right-hand side $\sfF$. In our applications, the system matrix $\sfA_0$ is given by \cref{eq:sys_matrix} without a complex shift, i.e.,~$\sfA_0 = \sfA(k, 0)$ and the corresponding right-hand side $\sfF$ is as in \cref{eq:discreteproblem}.
The preconditioner $\sfP_\varepsilon^{-1}$ is a single multigrid $V(\nu,\nu)$-cycle with $\nu \in \N$ pre- and postsmoothing steps applied to the preconditioner system matrix of the shifted problem $\sfA_\varepsilon = \sfA(k,\varepsilon)$.

Let $V_1 \subset V_{2} \ldots \subset V_{L} = V_h$ be a given hierarchy of $L \in \N$ discrete subspaces of $V_h$.
To each of these subspaces, we relate a preconditioner system matrix $\sfA_\varepsilon^{(\ell)}$ stemming from the discretization of \cref{eq:disc_var_form} with $V_\ell$ for $\ell = 1,\ldots,L$.
By $\sfI^{(\ell)}$, we denote the canonical interpolation or prolongation matrix, mapping a discrete function from $V_\ell$ to $V_{\ell+1}$ for $\ell = 1,\ldots,L-1$.
Moreover, we define a damped Jacobi smoother $\sfS_\varepsilon^{(\ell)}$ with damping factor $\omega = \frac{2}{3}$ for each $\sfA_\varepsilon^{(\ell)}$ with $\ell = 2,\ldots,L$.

The application of $\sfP_\varepsilon^{-1}$ is implemented by performing a $V(\nu,\nu)$-cycle to approximately solve the system $\sfP_\varepsilon \sfu = \sff$ for $\sfu$, given some right-hand side vector $\sff$.
This is achieved by calling the function \textsc{V-cycle}$(\sfu, \sff)$ given in \cref{alg:multigrid}.
\begin{algorithm}[H]
\caption{\label{alg:multigrid}Multigrid $V(\nu,\nu)$-cycle}
\begin{algorithmic}[1]
\Function{V-cycle}{$\sfu^{(\ell)}$, $\sfF^{(\ell)}$}
\If{$\ell = 1$}
\State\label{alg:line:coarsesolve}Solve $\sfA_\varepsilon^{(\ell)} \sfu^{(\ell)} = \sfF^{(\ell)}$ \Comment Direct solve
\Else
\State For $\nu$ steps $\sfu^{(\ell)} \gets \sfu^{(\ell)} + {\sfS_\varepsilon^{(\ell)}}^{-1}(\sfF^{(\ell)} - A_\varepsilon^{(\ell)} \sfu^{(\ell)})$ \Comment Presmoothing
\State $\sfr^{(\ell-1)} \gets {\sfI^{(\ell-1)}}^\T (\sfF^{(\ell)} - \sfA_\varepsilon^{(\ell)} \sfu^{(\ell)})$ \Comment Restrict residual
\State $\sfu^{(\ell-1)} \gets \bm{0}$ 
\State $\sfu^{(\ell-1)} \gets$ \Call{V-cycle}{$\sfu^{(\ell-1)}$, $\sfr^{(\ell-1)}$} \Comment Recursive coarse grid correction
\State $\sfu^{(\ell)} \gets \sfu^{(\ell)} + \sfI^{(\ell-1)} \sfu^{(\ell-1)}$ \Comment Add coarse grid correction
\State For $\nu$ steps $\sfu^{(\ell)} \gets \sfu^{(\ell)} + {\sfS^{(\ell)}}^{-\T}(\sfF^{(\ell)} - A_e^{(\ell)} \sfu^{(\ell)})$ \Comment Postsmoothing
\EndIf
\State \Return $\sfu^{(\ell)}$
\EndFunction
\end{algorithmic} 
\end{algorithm}
An important aspect of using this iterative method is that on the finer levels $\ell = 2,\ldots,L$, the matrices $\sfA_\varepsilon^{(\ell)}$, $\sfS_\varepsilon^{(\ell)}$, and $\sfI^{(\ell-1)}$ never need to be formed explicitly.
Only the result of their action to a vector is required.
More precisely, the action of $\sfA$ is implemented in a matrix-free fashion by summing the results of the underlying matrices $\sfK$ and $\sfM(k,\varepsilon)$.
Due to software limitations, the boundary matrix corresponding to $\sfB(k)$ was not implemented in a matrix-free way, but is assembled instead.
However, since the integration in the sesquilinearform of $b(\cdot,\cdot;k)$ is performed on the boundary, the number of non-zeros in $B(k)$ is of lower order $\mcO(h^{-d+1})$ compared to the other terms with $\mcO(h^{-d})$ non-zeros.

Using such a semi matrix-free approach effectively saves the memory required for storing the global sparse matrices which in turn allows for solving problems for which the matrices and its factorizations do not fit in memory.
Additionally, the direct solve in line \ref{alg:line:coarsesolve} of \cref{alg:multigrid} requires fewer resources.
Note that the smaller memory consumption is not the only advantage of a matrix-free approach, since it also significantly reduces the memory traffic.
It can be shown that using matrix-free methods instead of matrix-based approaches is beneficial for the performance and may even outperform matrix-vector multiplications with stored matrices \cite{Kronbichler2019Multigrid,drzisga2020stencil}.

In the remainder of this article, we restrict ourselves to a hierarchy of only two levels $L=2$ with spaces $V_2$ and $V_1 \subset V_2$.
Here, $V_2$ is formed by uniformly subdividing each element in the mesh corresponding to $V_1$.
The mesh size corresponding to $V_1$ is therefore $2h$.
Additionally, we do not consider any restarts in the FGMRES method.
\section{Data generation and processing}
\label{sec:data_generation}
\begin{figure}
\centering
\includegraphics[width=0.29850746268656714\textwidth]{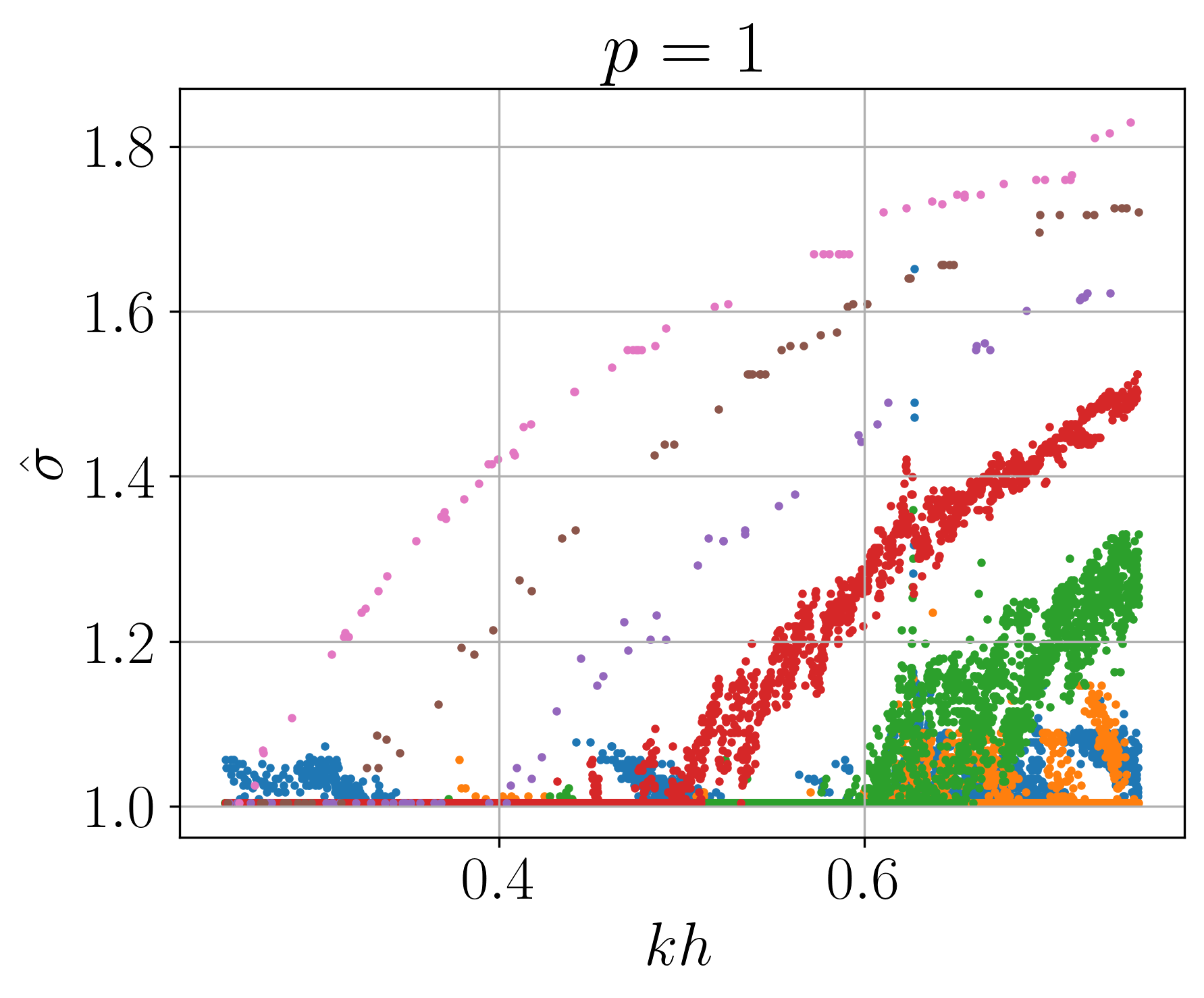}\includegraphics[width=0.29850746268656714\textwidth]{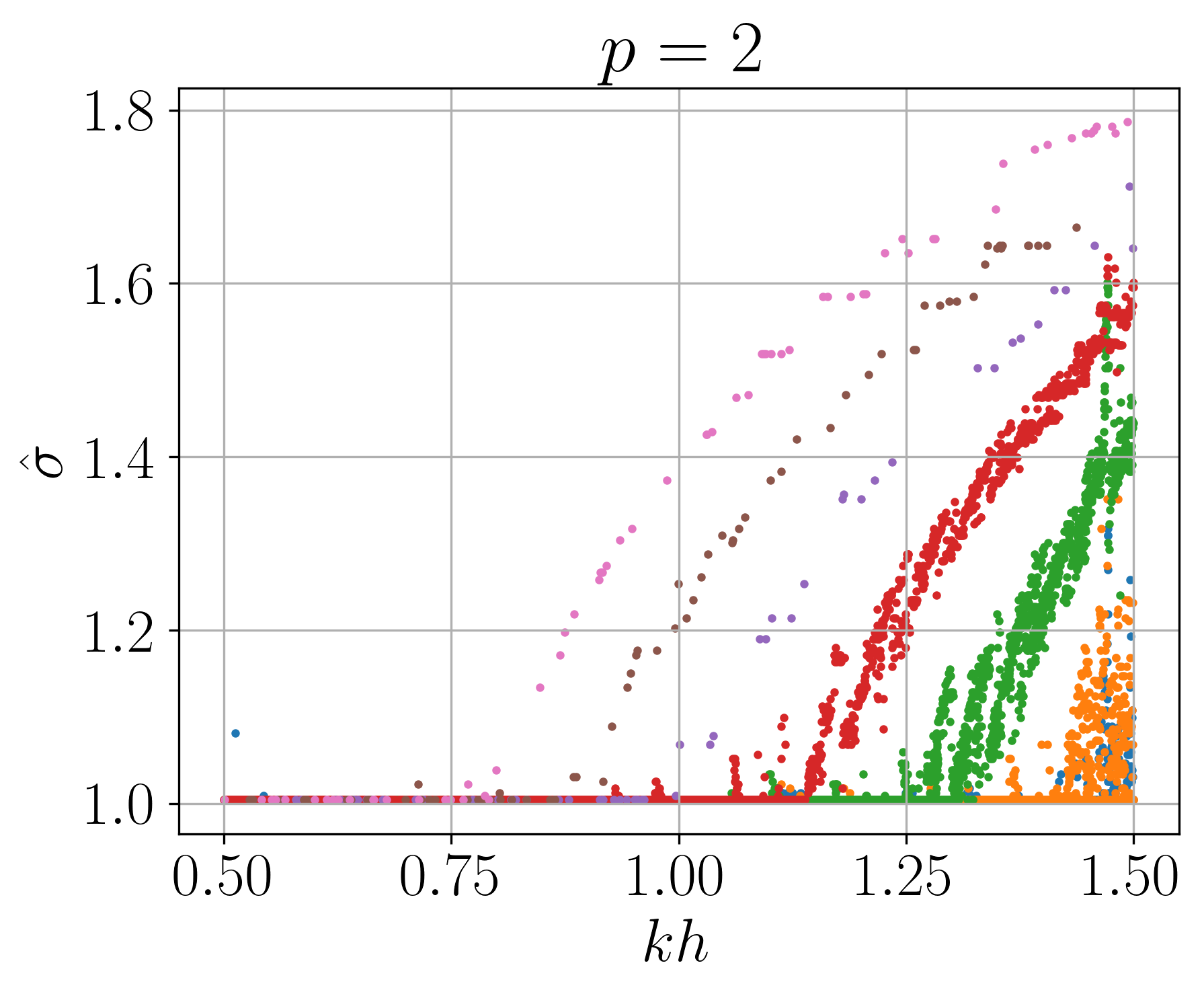}\includegraphics[width=0.29850746268656714\textwidth]{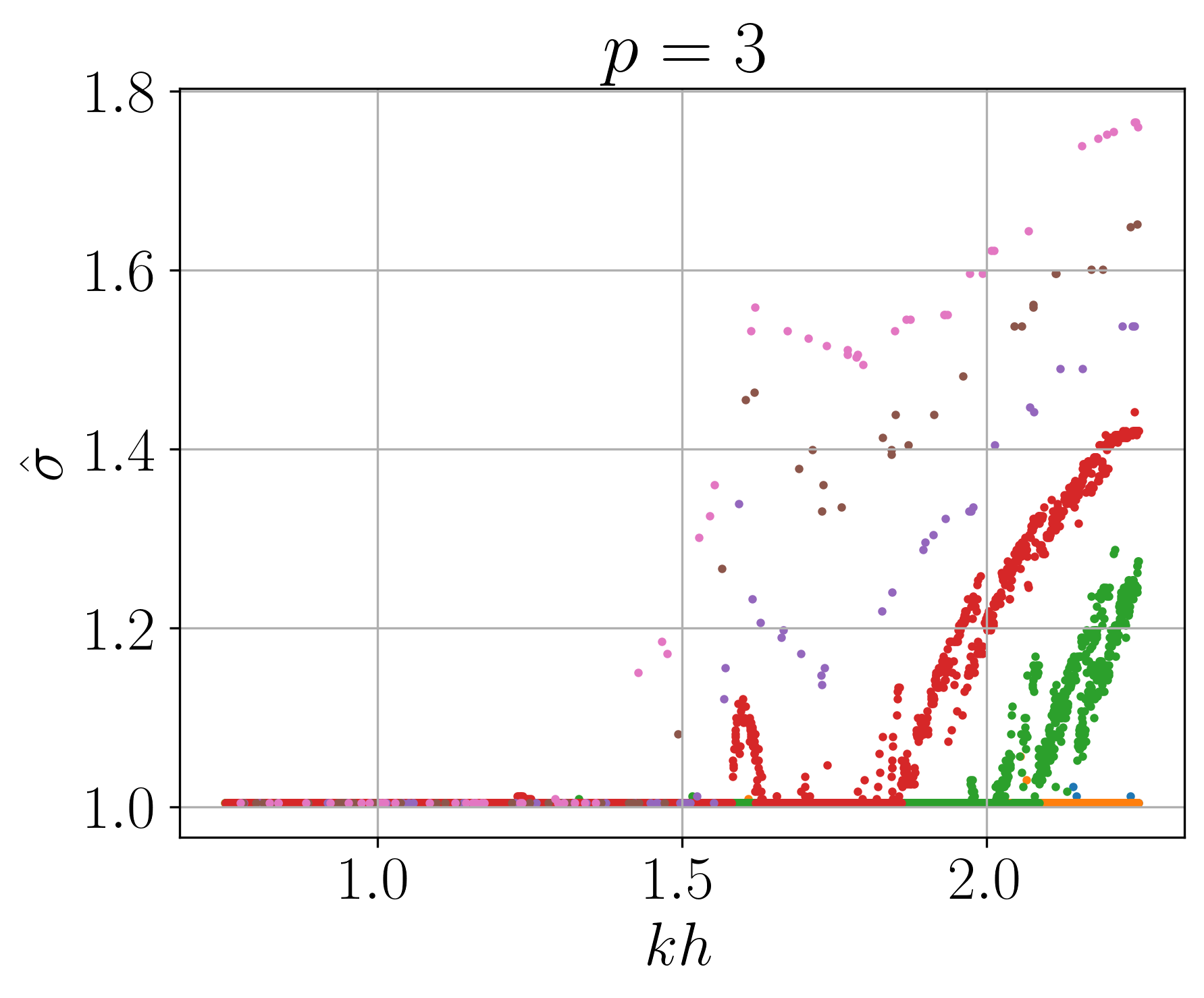}\includegraphics[width=0.10447761194029849\textwidth]{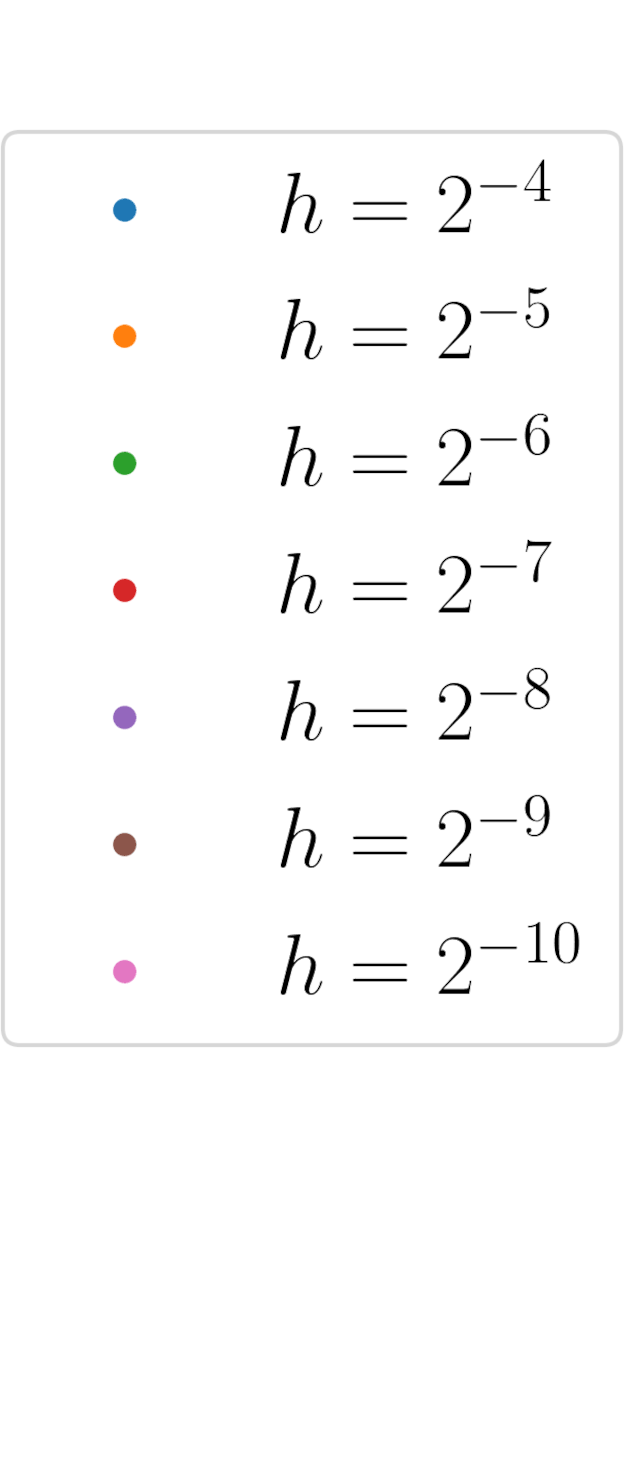}
\caption{\label{fig:OptimalMaps} Optimized shift exponents $\hat{\shiftexp}$ for decreasing $h$ and $p \in \{1,2,3\}$ from left to right.}
\end{figure}

\begin{figure}
\centering
\includegraphics[width=0.29850746268656714\textwidth]{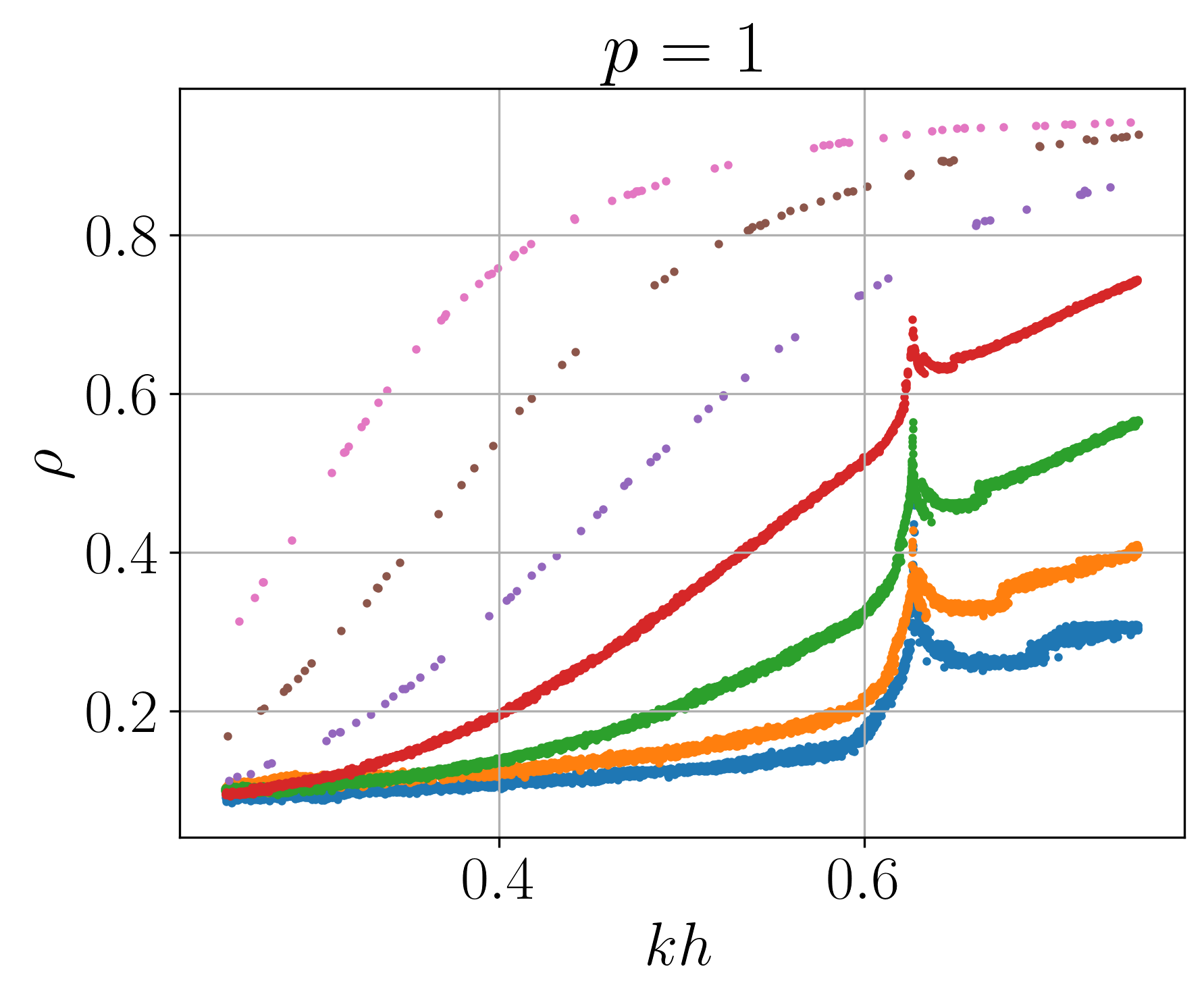}\includegraphics[width=0.29850746268656714\textwidth]{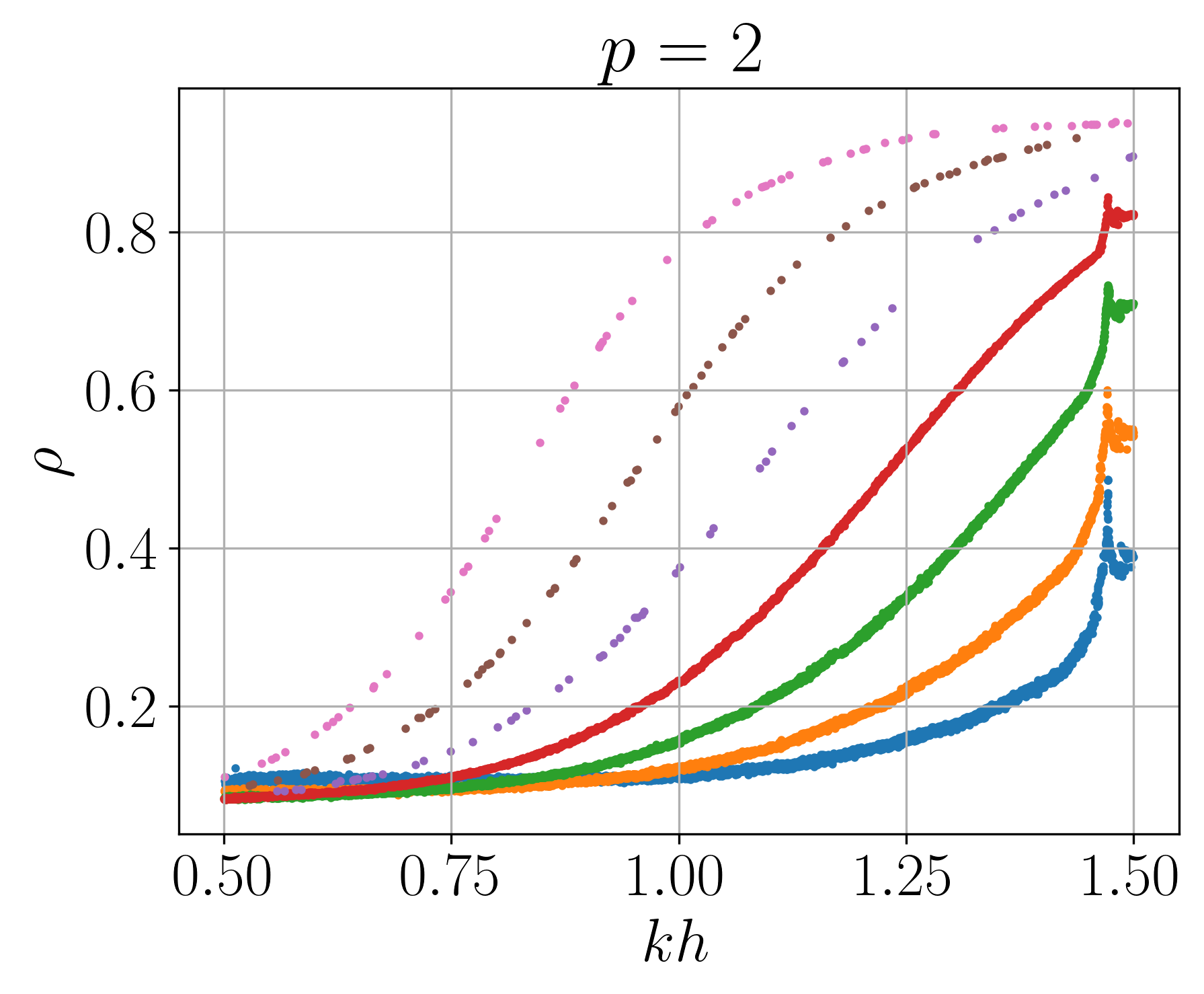}\includegraphics[width=0.29850746268656714\textwidth]{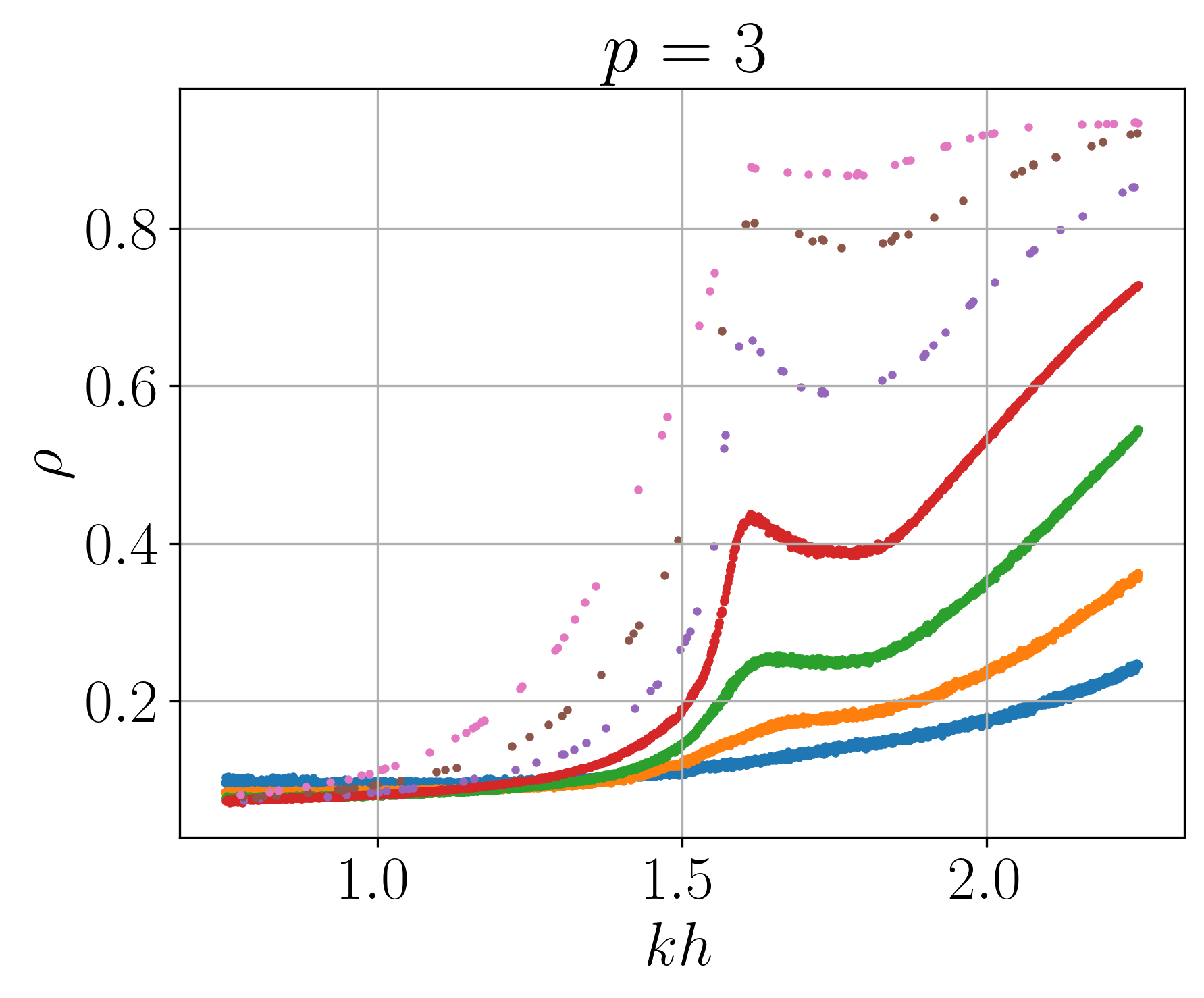}\includegraphics[width=0.10447761194029849\textwidth]{postprocess_legend.png}
\caption{\label{fig:OptimalRho} Average convergence rate $\rho$ using the optimized exponential shift for decreasing $h$ and $p \in \{1,2,3\}$ from left to right.}
\end{figure}

In this section, we present our approach to obtain the near optimal complex shift exponents $\hat{\shiftexp}$ depending on the wavenumber $k$ and discretization parameters $\ell \in \N$ with $h = 2^{-\ell}$ and $p \in \{1,2,3\}$.
The analysis in \Cref{sec:LFA} justifies the existence of a near optimal complex shift $\shiftexp$ for which the twogrid method converges, but this does not necessarily need to be optimal shift for which the FGMRES converges faster.
The ultimate goal is to construct a map which maps the input parameters to the near optimal shift exponent $\hat{\shiftexp}$, i.e.,~$(k,\ell,p) \mapsto \hat{\shiftexp}$.
In order to compute this map, we consider a data based approach in which we generate a set of samples $(k,\ell,p,\hat{\shiftexp})$ that are used in a subsequent nonlinear regression step. The following steps describes the process of generating such a single data point:\vspace*{1em}
\begin{enumerate}
\item Choose an order $p$ from the set $\{1,2,3\}$.
\item Choose a mesh size $h = 2^{-\ell}$ by selecting an $\ell$ from the set $\{4,\ldots,10\}$.
\item Choose a wavenumber from the interval $[\frac{3p}{16h}, \frac{3p}{4h}]$.
\item Generate a right-hand side vector $\sff$ with components laying uniformly in the interval $[-1,1]$.
\item Find the optimal exponent $\hat{\shiftexp} \in [1,2]$ of the complex shift $k^{\hat{\shiftexp}}$ using the gradient free golden-section search \cite{Kiefer1953Sequential} with a tolerance of $10^{-2}$. This involves repeatedly solving \cref{eq:Helmholtz} on $\Omega = (0,s)^2$ with $s = 1$ and $g = 0$ using the parameters $(k,\ell,p)$ and $\sff$.
\item Store the tuple $(k,\ell,p,\hat{\shiftexp})$ and go to Step 1.
\end{enumerate}\vspace*{1em}
Let $r_0$ be the residual norm at beginning of a solve in Step 5, i.e.,~$r_0 = \|\sfA_0\sfu^{(0)} - \sff\|_2$.
In this process, we solve until either a relative residual reduction of $10^{-8}$ is obtained or a maximum of $50$ iterations is reached.
Let $r_{\mathrm{end}} = \|\sfA_0\sfu^{(\mathrm{end})} - \sff\|_2$ be the final residual at termination after $\mathrm{end}$ iterations.
The objective function for the minimization in Step $5$ is then given by $\rho = (\frac{r_{\mathrm{end}}}{r_0})^{\frac{1}{\mathrm{end}}}$.
Note that the actual residual norm and not the residual norm of the preconditioned system is used.
Repeating the above process $N$ times will generate a set of training data $\mcD = \{(k_i,\ell_i,p_i,\hat{\shiftexp}_i)\}_{i=1}^{N}$.
Plots of the sample points for this particular case are presented in \cref{fig:OptimalMaps} for different values of $p$.
The corresponding average convergence rate $\rho$ for the near optimal complex shifts are illustrated in \cref{fig:OptimalRho}.
The idea of this approach is to approximate this map by a smooth function which can be evaluated for any reasonable input $(k,\ell,p)$.
Even if this data is obtained only for a constant wavenumber over the whole domain, we will use the approximated map to obtain optimal complex shifts for inhomogeneous wavenumbers in the numerical experiments in \Cref{sec:results}.

\begin{figure}
\centering
\includegraphics[width=0.33\textwidth]{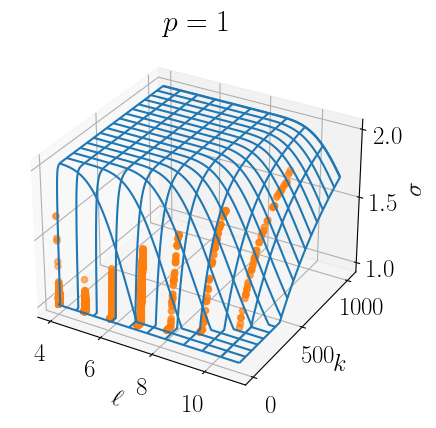}\includegraphics[width=0.33\textwidth]{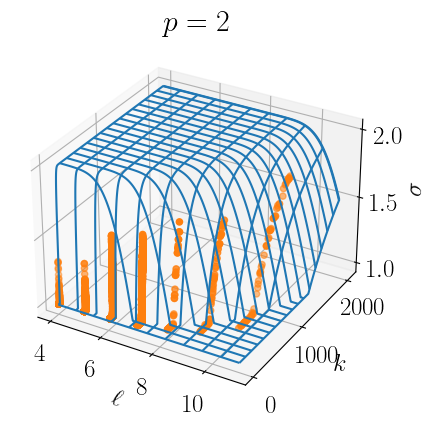}\includegraphics[width=0.33\textwidth]{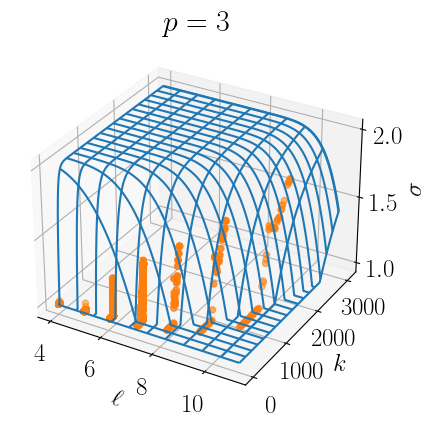}
\caption{\label{fig:Approximation}Sampling points in orange and approximated optimal shift map $\shiftexp_p$ in blue for $p \in \{1,2,3\}$ from left to right.}
\end{figure}

Let $N_{\ell,p} = |\{(k_i,\ell_i,p_i,\hat{\shiftexp}_i) \in \mcD \colon p_i = p\}|$ and $w_{\ell,p} = \frac{N_{\ell,p}}{|\mcD|}$.
For a vector of coefficients $(\hat{k}_{c,0}, \hat{k}_{c,1}, \hat{\alpha}_0, \hat{\alpha}_1)^\T \in \R^4$, let the approximated map $\shiftexp_p$ be defined as
\begin{align}
k_c(\ell) &= \hat{k}_{c,1} \cdot \exp(\hat{k}_{c,0} \cdot \ell),\\
\alpha(\ell) &= \hat{\alpha}_1 \cdot \exp(\hat{\alpha}_0 \cdot \ell),\\
\beta(k,\ell) &= 2 - \exp(-\alpha(\ell) \cdot (k - k_c(\ell))),\\
\shiftexp_p(k,\ell) &= \min(\max(\beta(k,\ell), 1), 2).
\label{eq:shiftexp}
\end{align}
The objective function of our regression for a fixed $p$ is defined as
$$\mathrm{loss}_p(\hat{k}_{c,0}, \hat{k}_{c,1}, \hat{\alpha}_0, \hat{\alpha}_1) = \sum_{(k_i,\ell_i,p_i,\hat{\shiftexp}_i) \in \mcD, p_i = p} w_{\ell_i,p}^{-2} \cdot (\hat{\shiftexp}_{i} - \shiftexp_p(k_i, \ell_i))^2.$$
For each $p \in \{1,2,3\}$, we train a map $\shiftexp_p$ with the data obtained in the previous step.
This is achieved by optimizing for the vector of coefficients $(\hat{k}_{c,0}, \hat{k}_{c,1}, \hat{\alpha}_0, \hat{\alpha}_1)^\T \in \R^4$.
For this purpose, we use \texttt{PyTorch} \cite{NEURIPS2019_9015} together with the ADAM optimizer and a learning rate of $10^{-3}$.
The weights are initialized with $(0.1, 1.0, -0.5, 1.0)^\T$ and the final weights after $50\,000$ epochs are depicted in \cref{tab:optweights} for each $p$.
The approximated optimal shift maps together with the sampling points are illustrated in \cref{fig:Approximation}.
These optimized values are used in the MFEM \cite{mfem} solver code described in the next section.
\begin{table}
\centering
\small
\caption{\label{tab:optweights}Optimal weights in function $\shiftexp_p$ for $p \in \{1,2,3\}$.}
\begin{tabular}{c|c|c|c|c}
$p$ & $\hat{k}_{c,0}$ & $\hat{k}_{c,1}$ & $\hat{\alpha}_0$ & $\hat{\alpha}_1$ \\
\hline
1 & 0.4592788619853418 & 2.5790032999702346 & -0.6261637288068426 & 1.7580549857142198\\
2 & 0.5736926870738827 & 2.5729974893966001 & -0.6615199737374460 & 1.5966386518185063\\
3 & 0.6305770719029798 & 2.4284320222555804 & -0.4465407372367102 & 0.1287828338493968
\end{tabular}
\end{table}

Next, we compare the minimal exponents $\sigma_c$ from \Cref{sec:LFA} to the optimal shifts for the FGMRES solver.
For this purpose, we gradually enlarge the computational domain in order to exclude the influence of the boundary conditions.
By means of this comparison, we want to theoretically support the choice of the shift $k^\sigma$ obtained by the optimization procedure described in the first part of this section.

From the LFA, we obtain the shift exponent for which it is guaranteed that the twogrid method applied to the shifted Laplacian problem converges, provided that the domain is very large and boundary effects do not play a role.
This shift exponent, however, does not need to be the optimal one when used in conjunction with the outer FGMRES solver.
Since our data generation provides us with near optimal shift exponents of the whole solver, we can compare them to the values from the LFA.
We expect that, asymptotically, the near optimal shift exponents are larger than the LFA exponents, since these guarantee convergence.

\begin{figure}[H]
\centering
\includegraphics[width=0.29850746268656714\textwidth]{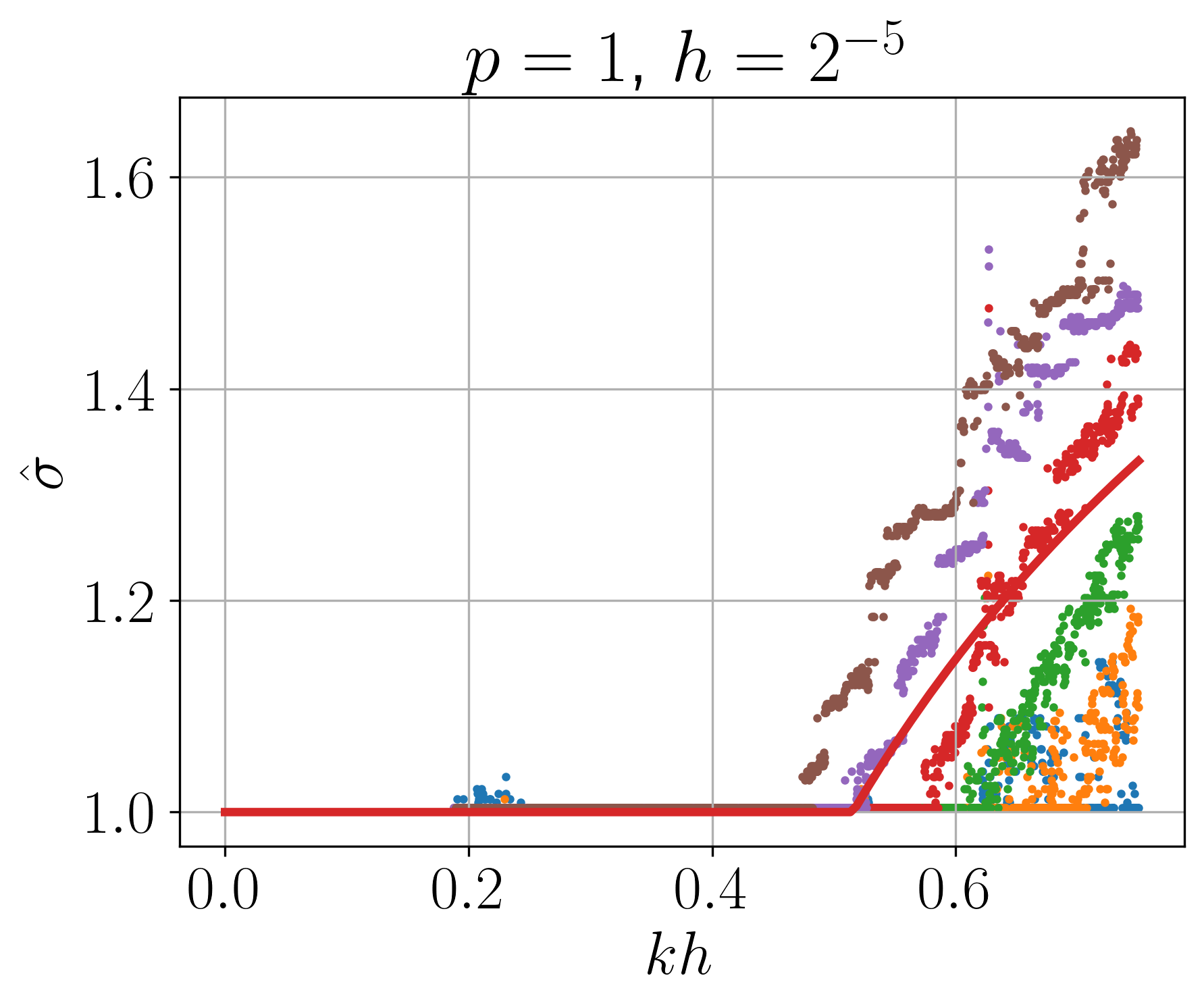}\includegraphics[width=0.29850746268656714\textwidth]{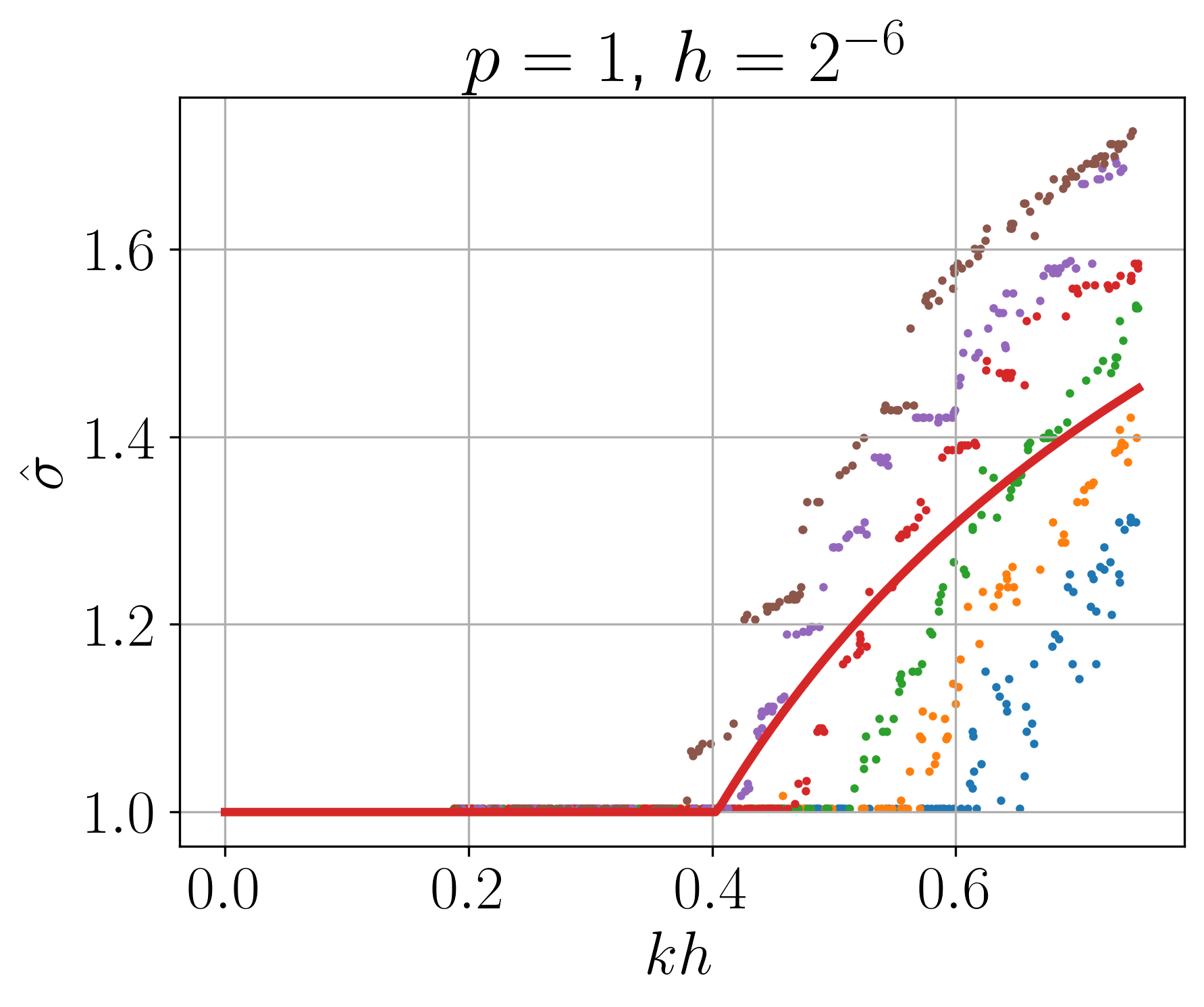}\includegraphics[width=0.29850746268656714\textwidth]{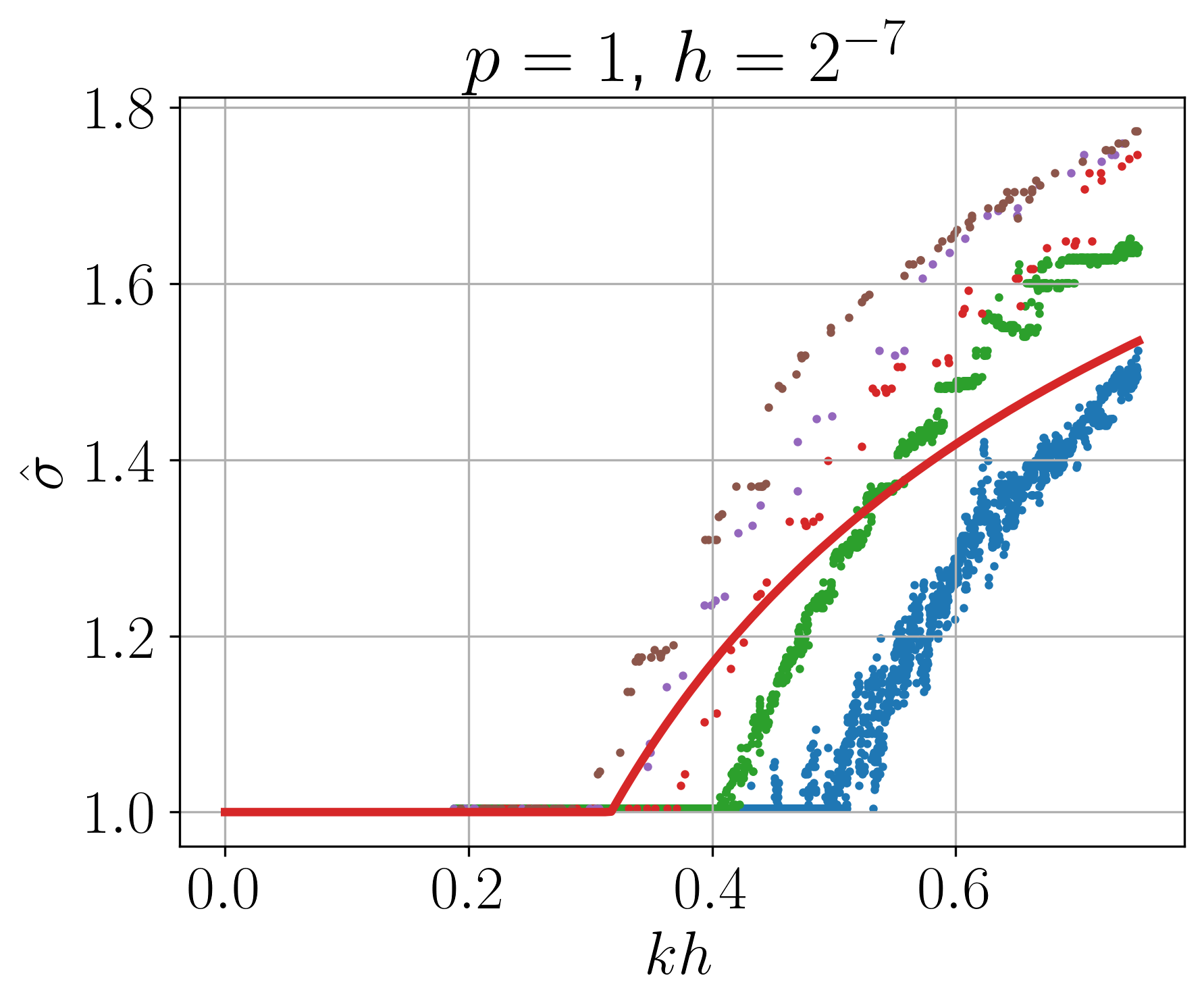}\includegraphics[width=0.10447761194029849\textwidth]{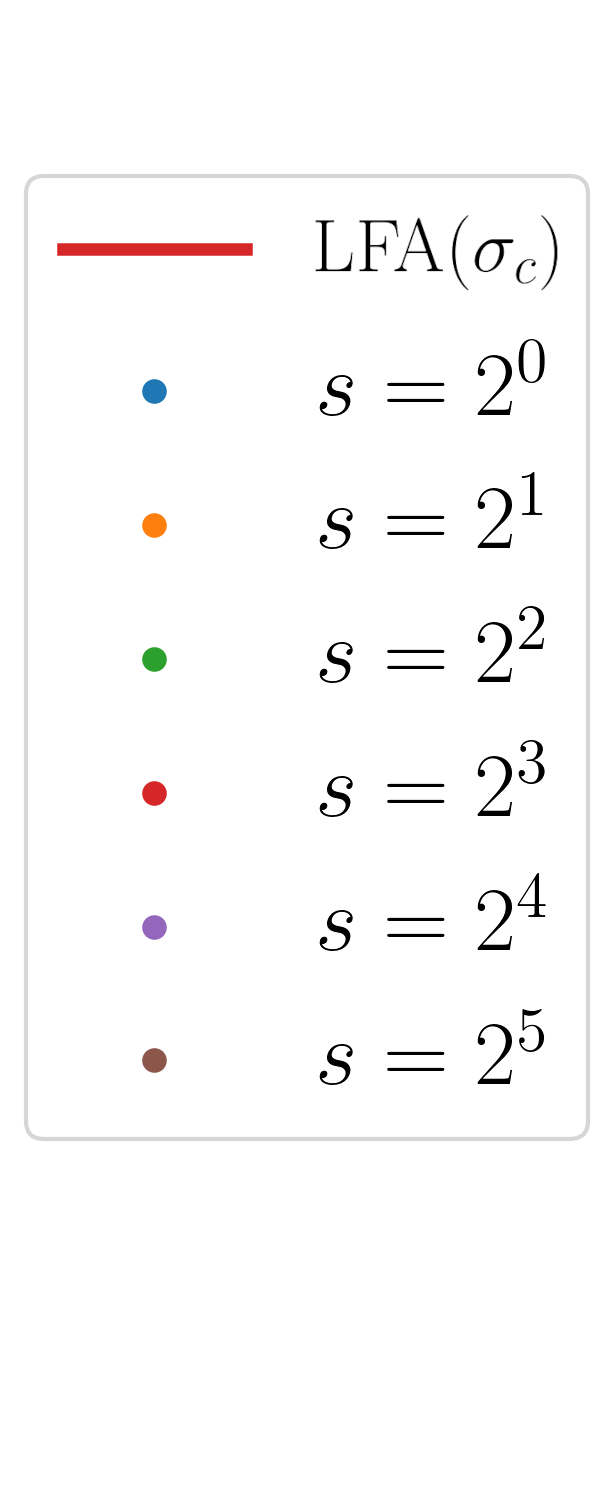}
\caption{\label{fig:lfa_comparison}Required shift exponent $\sigma_c$ obtained from the LFA and numerically sampled near optimal shifts on different domains $\Omega_s$ for $p=1$ and a fixed mesh size $h \in \{2^{-5}, 2^{-6}, 2^{-7}\}$.}
\end{figure}

The LFA yields results on an infinitely large domain, therefore, we need to make the near optimal shifts from the numerical experiments comparable.
For this purpose, we perform two different systematic comparisons.
In the first setup, we fix the parameters $p=1$ and set $h \in \{2^{-5}, 2^{-6}, 2^{-7}\}$.
In the second setup, we consider growing square domains $\Omega_s = (0, s)^2$ for $s \in \{2^i : i \in \{0,1,\ldots,5\}\}$ and keep the mesh size fixed.
For both setups, the number of degrees of freedom increases.
More precisely in \cref{fig:lfa_comparison}, we have an increase in the number of degrees of freedom within each picture from the blue to the brown color as well as from the left to the right picture within each color group.

We illustrate the results in \cref{fig:lfa_comparison} by plotting the required shift obtained from LFA alongside the near optimal shifts obtained by the data driven approach on different domains $\Omega_s$ and different fixed shifts $h$.
We may observe that the near optimal shift exponents are in fact larger than the required shift exponent, if the domain is enlarged.
\section{Numerical results}
\label{sec:results}
In this section, we demonstrate the effectiveness of using the approximated near optimal shift exponents when applied to solving \cref{eq:Helmholtz}.
For this purpose, we consider a set of synthetic and actual scenarios with heterogeneous wavenumbers.
In particular, we solve \cref{eq:Helmholtz} on $\Omega = (0,1)^d$, $d \in \{2,3\}$, with a source term
\begin{align}
f(\bmx) = 2 \cdot \exp(-1000 \cdot \|\bmx - \bms\|^2_2),
\end{align}
where the source location $\bms \in \R^d$ is specified in each of the scenarios. 
The additional boundary term $g$ is set to zero, i.e.,~$g = 0$.
In each of the following scenarios, we normalize the given velocity profile such that its values lie in the interval $[0,1]$ and denote this scaled profile by $\mu \colon \Omega \rightarrow [0,1]$.
Moreover, we choose a $k_{\mathrm{max}} \in \R$ and set the final heterogeneous wavenumber profile as $k(\bmx) = k_{\mathrm{max}} \cdot \mu(\bmx)$.

The solver with the two-grid preconditioner described in \Cref{sec:Multigrid} is implemented using the MFEM  modular finite element library \cite{mfem} with its support for multigrid operators.
The operators on the fine grid are implemented in a matrix-free fashion by using the partial assembly approach in which only the values at quadrature points are stored.
The operator applications reuse these precomputed data to compute the action of the operators on-the-fly.
The coarse grid problem is solved by computing the LU decomposition with MUMPS \cite{Amestoy2019Performance} through the PETSc \cite{petsc-web-page} interface within the first FGMRES iteration.
Subsequent iterations reuse the factorization for an efficient inversion of the coarse grid matrix.
The linear systems in each of the scenarios are solved without restarts and until a relative residual reduction $10^{-8}$ is achieved or the maximum number of 500 iterations is reached.
The data and source code for the nonlinear regression in \Cref{sec:data_generation} as well as the MFEM driver source code are available in the Zenodo archive\fnurl{https://doi.org/10.5281/zenodo.4607330}.

All run-time measurements for the 2D examples are obtained on a machine equipped with two Intel Xeon Gold 6136 processors with a nominal base frequency of 3.0 GHz.
Each processor has 12 physical cores which results in a total of 24 physical cores.
The total available memory of \SI{251}{\giga\byte} is split into two NUMA domains, one for each socket.
All available 24 physical cores are used for each computation.

The measurements for the 3D examples are conducted on the SuperMUC-NG system equipped with Skylake nodes.
The following values were taken from \cite{SuperMUCNG:Configuration:Details}.
Each node has two Intel Xeon Platinum 8174 processors with a nominal clock rate of 3.1 GHz.
Each processor has 24 physical cores which results in 48 cores per node.
Each core has a dedicated L1 (data) cache of size 32\,kB and a dedicated L2 cache of size 1024\,kB.
Each of the two processors has a L3 cache of size 33\,MB shared across all its cores.
The total main memory of 94\,GB is split into equal parts across two NUMA domains with one processor each.
We use the native Intel 19.0 compiler together with the Intel 2019 MPI library.

In the following subsections, we consider scenarios with velocity profiles stemming from different sources: an artificial wedge example and three profiles from synthetic and non-synthetic geological cross-sections.
Lastly, we consider a 3D scenario in which one of the 2D cross-sections are extruded to 3D.
For each possible scenario, we compare the outer FGMRES iteration numbers for different values of the complex shift $\varepsilon \in \{0, k, k^{\frac{3}{2}}, k^2, k^{\shiftexp_p}\}$.
By $k^\shiftexp$, we denote the case in which the optimal shift exponent map $\shiftexp_p$ from \cref{eq:shiftexp} is used with the respective coefficients from \cref{tab:optweights}.
The speed-ups are computed with respect to the case without a shift, i.e.,~$\varepsilon = 0$.

\subsection{Wedge example}
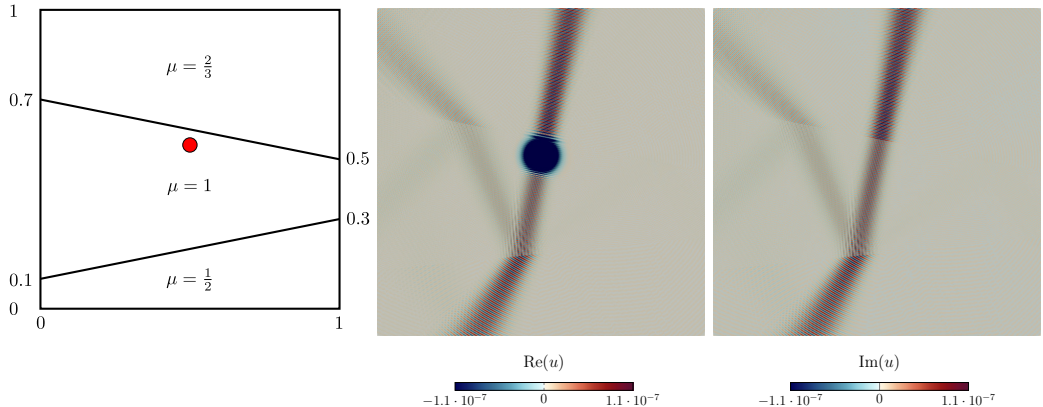
\begin{figure}[H]
\centering
\begin{minipage}{0.33\textwidth}
\begin{center}
\includegraphics[width=\textwidth]{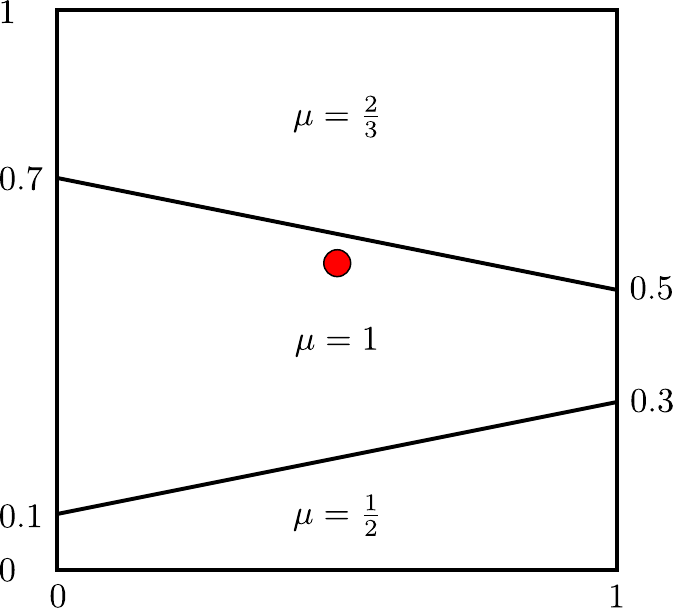}\\[0.5em]
\phantom{\begin{scaletikzpicturetowidth}{0.75\textwidth}
\begin{tikzpicture}[scale=\tikzscale,font=\large]
\pgfmathsetlengthmacro\MajorTickLength{
  \pgfkeysvalueof{/pgfplots/major tick length} * 0.5
}
\begin{axis}[
title={\Large $\mu$},
xmin=0, xmax=1,
ymin=0, ymax=0.02,
axis on top,
scaled x ticks=false,
scaled y ticks=false,
ytick=\empty,
yticklabels=\empty,
yticklabel pos=right,
x tick label style={
  /pgf/number format/.cd,
                fixed,
        precision=1,
  /tikz/.cd  
},
extra x tick style={
    font=\large,
    tick style=transparent,     yticklabel pos=left,
    x tick label style={
        /pgf/number format/.cd,
            std,
            precision=3,
      /tikz/.cd
    }
},
width=7cm,
height=1.82cm,
major tick length=\MajorTickLength,
max space between ticks=1000pt,
try min ticks=4,
]
\addplot graphics [
includegraphics cmd=\pgfimage,
xmin=\pgfkeysvalueof{/pgfplots/xmin}, 
xmax=\pgfkeysvalueof{/pgfplots/xmax}, 
ymin=\pgfkeysvalueof{/pgfplots/ymin}, 
ymax=\pgfkeysvalueof{/pgfplots/ymax}
] {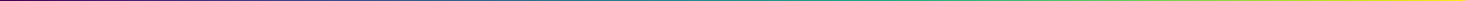};
\end{axis}
\end{tikzpicture}
\end{scaletikzpicturetowidth}}
\end{center}
\end{minipage}
\begin{minipage}{0.3\textwidth}
\begin{center}
\includegraphics[width=\textwidth]{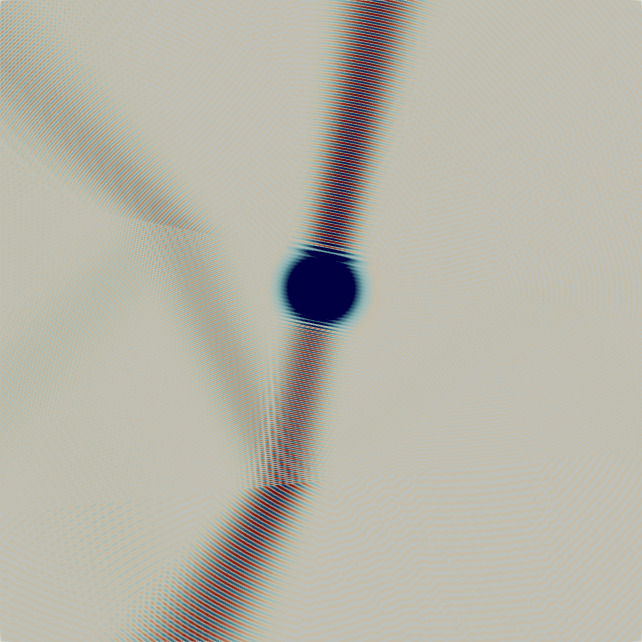}\\[0.5em]
\begin{scaletikzpicturetowidth}{0.75\textwidth}
\begin{tikzpicture}[scale=\tikzscale,font=\large]
\pgfmathsetlengthmacro\MajorTickLength{
  \pgfkeysvalueof{/pgfplots/major tick length} * 0.5
}
\begin{axis}[
title={\Large $\mathrm{Re}(u)$},
xmin=-1.1e-7, xmax=1.1e-7,
ymin=0, ymax=0.02,
axis on top,
scaled x ticks=false,
scaled y ticks=false,
ytick=\empty,
yticklabels=\empty,
yticklabel pos=right,
xtick = {0},
xticklabels = {0},
extra x ticks={
   \pgfkeysvalueof{/pgfplots/xmin},
   \pgfkeysvalueof{/pgfplots/xmax}
},
extra x tick style={
    font=\large,
    tick style=transparent,     yticklabel pos=left,
    x tick label style={
        /pgf/number format/.cd,
            std,
      /tikz/.cd
    }
},
width=7cm,
height=1.82cm,
major tick length=\MajorTickLength,
max space between ticks=1000pt,
try min ticks=2,
]
\addplot graphics [
includegraphics cmd=\pgfimage,
xmin=\pgfkeysvalueof{/pgfplots/xmin}, 
xmax=\pgfkeysvalueof{/pgfplots/xmax}, 
ymin=\pgfkeysvalueof{/pgfplots/ymin}, 
ymax=\pgfkeysvalueof{/pgfplots/ymax}
] {./figures/colormap_cool_to_warm_extended_rotated.png};
\end{axis}
\end{tikzpicture}
\end{scaletikzpicturetowidth}
\end{center}
\end{minipage}
\begin{minipage}{0.3\textwidth}
\begin{center}
\includegraphics[width=\textwidth]{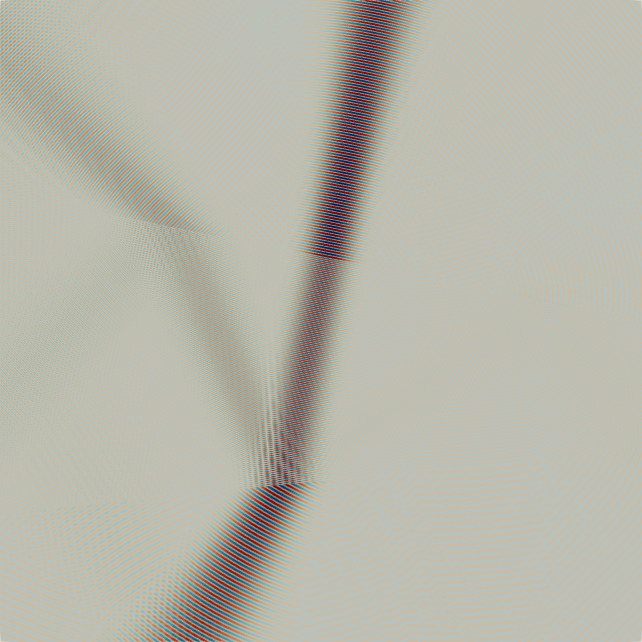}\\[0.5em]
\begin{scaletikzpicturetowidth}{0.75\textwidth}
\begin{tikzpicture}[scale=\tikzscale,font=\large]
\pgfmathsetlengthmacro\MajorTickLength{
  \pgfkeysvalueof{/pgfplots/major tick length} * 0.5
}
\begin{axis}[
title={\Large $\mathrm{Im}(u)$},
xmin=-1.1e-7, xmax=1.1e-7,
ymin=0, ymax=0.02,
axis on top,
scaled x ticks=false,
scaled y ticks=false,
ytick=\empty,
yticklabels=\empty,
yticklabel pos=right,
xtick = {0},
xticklabels = {0},
extra x ticks={
   \pgfkeysvalueof{/pgfplots/xmin},
   \pgfkeysvalueof{/pgfplots/xmax}
},
extra x tick style={
    font=\large,
    tick style=transparent,     yticklabel pos=left,
    x tick label style={
        /pgf/number format/.cd,
            std,
      /tikz/.cd
    }
},
width=7cm,
height=1.82cm,
]
\addplot graphics [
includegraphics cmd=\pgfimage,
xmin=\pgfkeysvalueof{/pgfplots/xmin}, 
xmax=\pgfkeysvalueof{/pgfplots/xmax}, 
ymin=\pgfkeysvalueof{/pgfplots/ymin}, 
ymax=\pgfkeysvalueof{/pgfplots/ymax}
] {./figures/colormap_cool_to_warm_extended_rotated.png};
\end{axis}
\end{tikzpicture}
\end{scaletikzpicturetowidth}
\end{center}
\end{minipage}
\caption{\label{fig:wedge2dfigures} Wedge example velocity profile with source location (left) and real and imaginary part of the solution in the case $p=2$ (center and right).}
\end{figure}

In the first scenario, we consider an artificial velocity profile~$\mu$ with three distinct values as illustrated in the left of \cref{fig:wedge2dfigures}.
The maximum value of the source term is located at $\bms = (0.5, 0.55)^\T$.
We collect the required number of FGMRES iterations and the respective compute times for $p \in \{1,2,3\}$ and the maximum wavenumbers $k_{\mathrm{max}} \in \{450, 1100, 1800\}$ in \cref{tab:wedge2d}.
Additionally, in the center and right of \cref{fig:wedge2dfigures}, we present the real and imaginary parts of the solution in the case of $p=2$.
We observe that using the near optimal shift exponent results in the smallest number of iterations throughout all the considered cases.
However, due to a faster LU factorization, using a shift $k^{\frac{3}{2}}$ still results in a shorter compute time for $p=2$ even if five more iterations are performed.

\begin{table}[H]
\centering
\footnotesize
\caption{\label{tab:wedge2d}Parameter values, iteration numbers, and compute times for the 2D wedge example.}
\begin{tabular}{ll}
\begin{tabular}[t]{l|l}
Param. & Value\\
\hline
$p$ & $1$\\
$h$ & $2^{-10}$\\
$k_{\mathrm{max}}$ & $450$\\
$k_{\mathrm{max}} \cdot h$ & $0.585938$\\
DoFs & $2\,101\,250$
\end{tabular}
&
\begin{tabular}[t]{c|r|r|r}
$\varepsilon$ & Iter & Time [m:s] & Speed-up \\
\hline
$0$               & 412 & 3:33.68 &\\
$k$               & 330 & 2:10.38 & 63.89\%\\
$k^{\frac{3}{2}}$ &  96 & 0:32.16 & 564.43\%\\
$k^{2}$           &>500 &      -- & --\\
$k^\shiftexp$     &  93 & 0:31.29 & 582.90\%
\end{tabular}\vspace*{0.5em}\\
\begin{tabular}[t]{l|l}
Param. & Value\\
\hline
$p$ & $2$\\
$h$ & $2^{-10}$\\
$k_{\mathrm{max}}$ & $1100$\\
$k_{\mathrm{max}} \cdot h$ & $1.07422$\\
DoFs & $8\,396\,802$
\end{tabular}
&
\begin{tabular}[t]{c|r|r|r}
$\varepsilon$ & Iter & Time [m:s] & Speed-up \\
\hline
$0$               & 491 & 19:01.23 &  \\
$k$               & 385 & 12:52.17 & 47.80\%\\
$k^{\frac{3}{2}}$ & 114 &  3:09.02 & 503.76\%\\
$k^{2}$           &>500 &       -- & --\\
$k^\shiftexp$     & 109 &  3:18.40 & 475.22\%
\end{tabular}\vspace*{0.5em}\\

\begin{tabular}[t]{l|l}
Param. & Value\\
\hline
$p$ & $3$\\
$h$ & $2^{-10}$\\
$k_{\mathrm{max}}$ & $1800$\\
$k_{\mathrm{max}} \cdot h$ & $1.75781$\\
DoFs & $18\,886\,658$
\end{tabular}
&
\begin{tabular}[t]{c|r|r|r}
$\varepsilon$ & Iter & Time [m:s] & Speed-up \\
\hline
$0$               & 371 & 28:04.96 & \\
$k$               & 297 & 21:04.68 & 33.23\%\\
$k^{\frac{3}{2}}$ & 111 & 7:13.79 & 288.43\%\\
$k^{2}$           &>500 &       -- &  --\\
$k^\shiftexp$     & 106 & 6:53.57 & 307.42\%
\end{tabular}
\end{tabular}
\end{table}

\subsection{Marmousi model}
\begin{figure}[H]
\centering
\begin{minipage}{0.32\textwidth}
\begin{center}
\includegraphics[width=\textwidth]{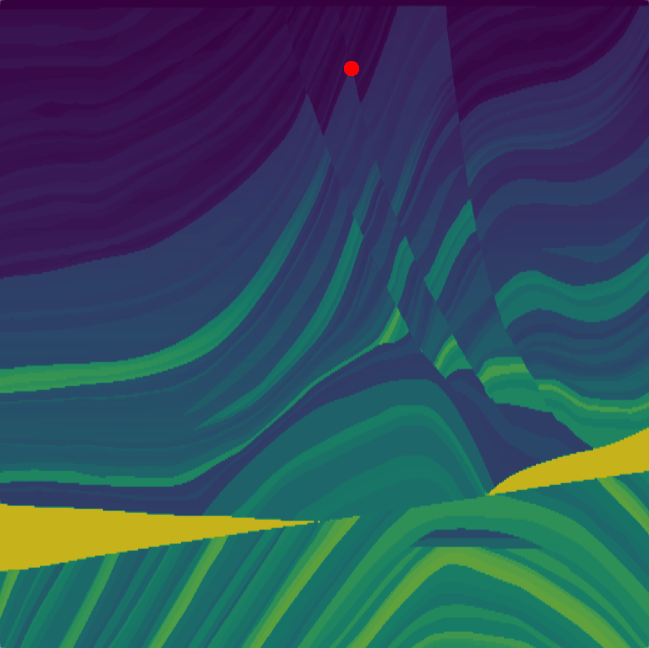}\\[0.5em]
\begin{scaletikzpicturetowidth}{0.75\textwidth}
\begin{tikzpicture}[scale=\tikzscale,font=\large]
\pgfmathsetlengthmacro\MajorTickLength{
  \pgfkeysvalueof{/pgfplots/major tick length} * 0.5
}
\begin{axis}[
title={\Large $\mu$},
xmin=0, xmax=1,
ymin=0, ymax=0.02,
axis on top,
scaled x ticks=false,
scaled y ticks=false,
ytick=\empty,
yticklabels=\empty,
yticklabel pos=right,
x tick label style={
  /pgf/number format/.cd,
                fixed,
        precision=1,
  /tikz/.cd  
},
extra x tick style={
    font=\large,
    tick style=transparent,     yticklabel pos=left,
    x tick label style={
        /pgf/number format/.cd,
            std,
            precision=3,
      /tikz/.cd
    }
},
width=7cm,
height=1.82cm,
major tick length=\MajorTickLength,
max space between ticks=1000pt,
try min ticks=4,
]
\addplot graphics [
includegraphics cmd=\pgfimage,
xmin=\pgfkeysvalueof{/pgfplots/xmin}, 
xmax=\pgfkeysvalueof{/pgfplots/xmax}, 
ymin=\pgfkeysvalueof{/pgfplots/ymin}, 
ymax=\pgfkeysvalueof{/pgfplots/ymax}
] {./figures/colormap_viridis_rotated.png};
\end{axis}
\end{tikzpicture}
\end{scaletikzpicturetowidth}
\end{center}
\end{minipage}
\begin{minipage}{0.32\textwidth}
\begin{center}
\includegraphics[width=\textwidth]{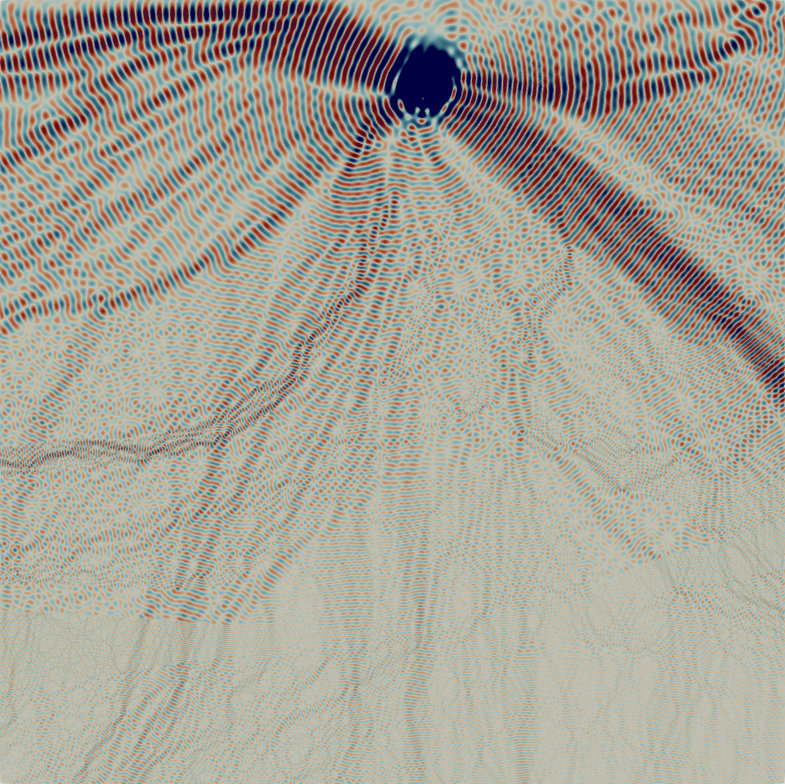}\\[0.5em]
\begin{scaletikzpicturetowidth}{0.75\textwidth}
\begin{tikzpicture}[scale=\tikzscale,font=\large]
\pgfmathsetlengthmacro\MajorTickLength{
  \pgfkeysvalueof{/pgfplots/major tick length} * 0.5
}
\begin{axis}[
title={\Large $\mathrm{Re}(u)$},
xmin=-2e-6, xmax=2e-6,
ymin=0, ymax=0.02,
axis on top,
scaled x ticks=false,
scaled y ticks=false,
ytick=\empty,
yticklabels=\empty,
yticklabel pos=right,
width=7cm,
height=1.82cm,
major tick length=\MajorTickLength,
max space between ticks=1000pt,
try min ticks=2,
]
\addplot graphics [
includegraphics cmd=\pgfimage,
xmin=\pgfkeysvalueof{/pgfplots/xmin}, 
xmax=\pgfkeysvalueof{/pgfplots/xmax}, 
ymin=\pgfkeysvalueof{/pgfplots/ymin}, 
ymax=\pgfkeysvalueof{/pgfplots/ymax}
] {./figures/colormap_cool_to_warm_extended_rotated.png};
\end{axis}
\end{tikzpicture}
\end{scaletikzpicturetowidth}
\end{center}
\end{minipage}
\begin{minipage}{0.32\textwidth}
\begin{center}
\includegraphics[width=\textwidth]{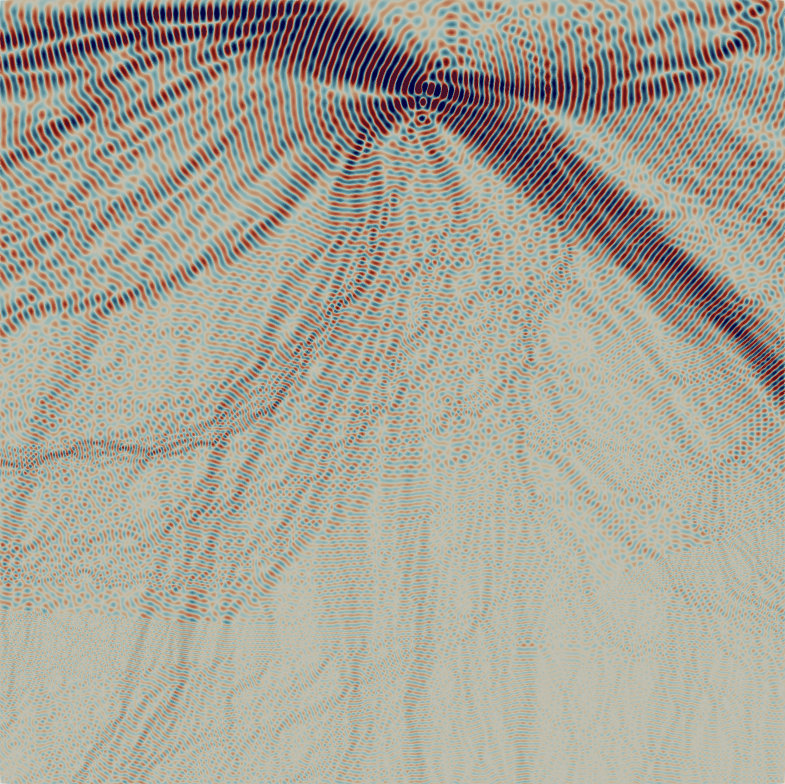}\\[0.5em]
\begin{scaletikzpicturetowidth}{0.75\textwidth}
\begin{tikzpicture}[scale=\tikzscale,font=\large]
\pgfmathsetlengthmacro\MajorTickLength{
  \pgfkeysvalueof{/pgfplots/major tick length} * 0.5
}
\begin{axis}[
title={\Large $\mathrm{Im}(u)$},
xmin=-2e-6, xmax=2e-6,
ymin=0, ymax=0.02,
axis on top,
scaled x ticks=false,
scaled y ticks=false,
ytick=\empty,
yticklabels=\empty,
yticklabel pos=right,
extra x ticks={
   \pgfkeysvalueof{/pgfplots/xmin},
   \pgfkeysvalueof{/pgfplots/xmax}
},
width=7cm,
height=1.82cm,
major tick length=\MajorTickLength,
max space between ticks=1000pt,
try min ticks=2,
]
\addplot graphics [
includegraphics cmd=\pgfimage,
xmin=\pgfkeysvalueof{/pgfplots/xmin}, 
xmax=\pgfkeysvalueof{/pgfplots/xmax}, 
ymin=\pgfkeysvalueof{/pgfplots/ymin}, 
ymax=\pgfkeysvalueof{/pgfplots/ymax}
] {./figures/colormap_cool_to_warm_extended_rotated.png};
\end{axis}
\end{tikzpicture}
\end{scaletikzpicturetowidth}
\end{center}
\end{minipage}
\caption{\label{fig:marmousi2dfigures} Marmousi example velocity profile with source location (left) and real and imaginary part of the solution in the case $p=2$ (center and right).}
\end{figure}
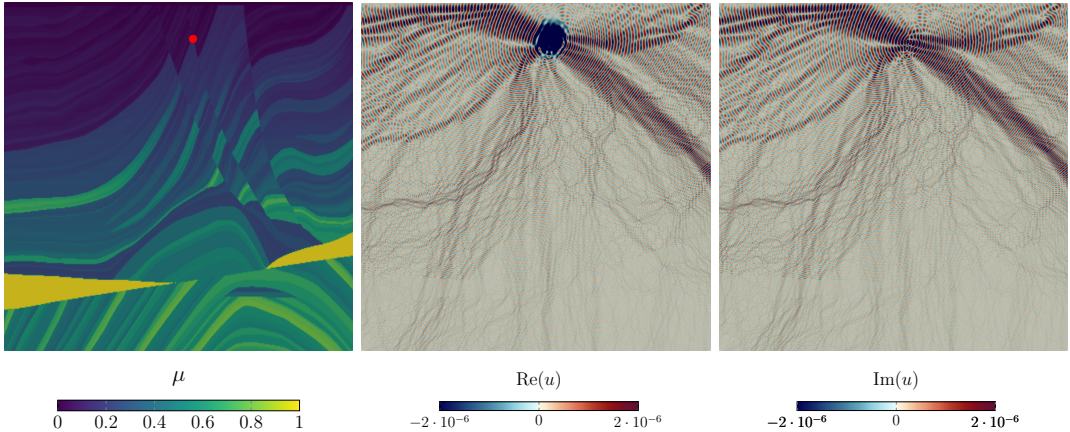
In a second scenario, we consider the velocity profile stemming from the synthetic Marmousi model devised by the Institut Fran\c{c}ais du Petrole \cite{Versteeg1994Marmousi}.
The corresponding scaled velocity profile is illustrated in the left of \cref{fig:marmousi2dfigures}.
Here, the maximum value of the source term is located at $\bms = (0.5421, 0.8946)^\T$.
We collect the required number of FGMRES iterations and the respective compute times for $p \in \{1,2,3\}$ and the maximum wavenumbers $k_{\mathrm{max}} \in \{600, 800, 1250, 1900\}$ in \cref{tab:marmousi2d}.
Additionally, in the center and right of \cref{fig:marmousi2dfigures}, we present the real and imaginary parts of the solution in the case of $p=2$.
We observe that using the near optimal shift exponent results in the smallest number of iterations for large wavenumbers.
For a smaller wavenumber $k_{\mathrm{max}} = 800$ and $p=2$, the trained shift still yields the minimal number of required iterations when compared to the other shifts.
This means that using the trained shift does not worsen the performance if the wavenumber is not in the critical regime. 
Using a shift $k^{\frac{3}{2}}$ results in a slightly shorter compute time for $p=1$ even if eight more iterations are performed, since the LU factorization required more time.
This cannot observed for the higher order cases with $p > 1$.

\begin{table}[H]
\centering
\footnotesize
\caption{\label{tab:marmousi2d}Parameter values, iteration numbers, and compute times for the 2D Marmousi example.}
\begin{tabular}{ll}
\begin{tabular}[t]{l|l}
Param. & Value\\
\hline
$p$ & $1$\\
$h$ & $2^{-10}$\\
$k_{\mathrm{max}}$ & $600$\\
$k_{\mathrm{max}} \cdot h$ & $0.5859$\\
DoFs & $2\,101\,250$
\end{tabular}
&
\begin{tabular}[t]{c|r|r|r}
$\varepsilon$ & Iter & Time [m:s] & Speed-up \\
\hline
$0$               & 358 & 3:03.67  &  \\
$k$               & 312 & 1:59.62 & 53.54\% \\
$k^{\frac{3}{2}}$ & 147 & 0:49.81 & 268.74\% \\
$k^{2}$           &>500 & --      & -- \\
$k^\shiftexp$     & 139 & 0:50.73 & 262.05\%
\end{tabular}\vspace*{0.5em}\\

\begin{tabular}[t]{l|l}
Param. & Value\\
\hline
$p$ & $2$\\
$h$ & $2^{-10}$\\
$k_{\mathrm{max}}$ & $800$\\
$k_{\mathrm{max}} \cdot h$ & $0.78125$\\
DoFs & $8\,396\,802$
\end{tabular}
&
\begin{tabular}[t]{c|r|r|r}
$\varepsilon$ & Iter & Time [m:s] & Speed-up \\
\hline
$0$               & 15  & 0:55.37 &  \\
$k$               & 15  & 0:41.30 & 34.07\% \\
$k^{\frac{3}{2}}$ & 55  & 1:47.19 & -48.34\% \\
$k^{2}$           &>500 & --      & -- \\
$k^\shiftexp$     & 15  & 0:41.41 & 33.71\%
\end{tabular}\vspace*{0.5em}\\

\begin{tabular}[t]{l|l}
Param. & Value\\
\hline
$p$ & $2$\\
$h$ & $2^{-10}$\\
$k_{\mathrm{max}}$ & $1250$\\
$k_{\mathrm{max}} \cdot h$ & $1.2207$\\
DoFs & $8\,396\,802$
\end{tabular}
&
\begin{tabular}[t]{c|r|r|r}
$\varepsilon$ & Iter & Time [m:s] & Speed-up \\
\hline
$0$               & 197 & 6:54.95 &  \\
$k$               & 181 & 5:33.39 & 24.46\% \\
$k^{\frac{3}{2}}$ & 125 & 3:44.19 & 85.09\% \\
$k^{2}$           &>500 & --      & -- \\
$k^\shiftexp$     & 109 & 3:21.13 & 106.31\%
\end{tabular}\vspace*{0.5em}\\

\begin{tabular}[t]{l|l}
Param. & Value\\
\hline
$p$ & $3$\\
$h$ & $2^{-10}$\\
$k_{\mathrm{max}}$ & $1900$\\
$k_{\mathrm{max}} \cdot h$ & $1.85547$\\
DoFs & $18\,886\,658$
\end{tabular}
&
\begin{tabular}[t]{c|r|r|r}
$\varepsilon$ & Iter & Time [m:s] & Speed-up \\
\hline
$0$               & 89   & 6:08.71 &  \\
$k$               & 85   & 5:35.00 & 10.06\% \\
$k^{\frac{3}{2}}$ & 99   & 6:29.95 & -5.45\% \\
$k^{2}$           & >500 & --      & -- \\
$k^\shiftexp$     & 76   & 5:01.33 & 22.36\%
\end{tabular}
\end{tabular}
\end{table}

\subsection{Migration from topography}
\begin{figure}[H]
\centering
\begin{minipage}{0.32\textwidth}
\begin{center}
\includegraphics[width=\textwidth]{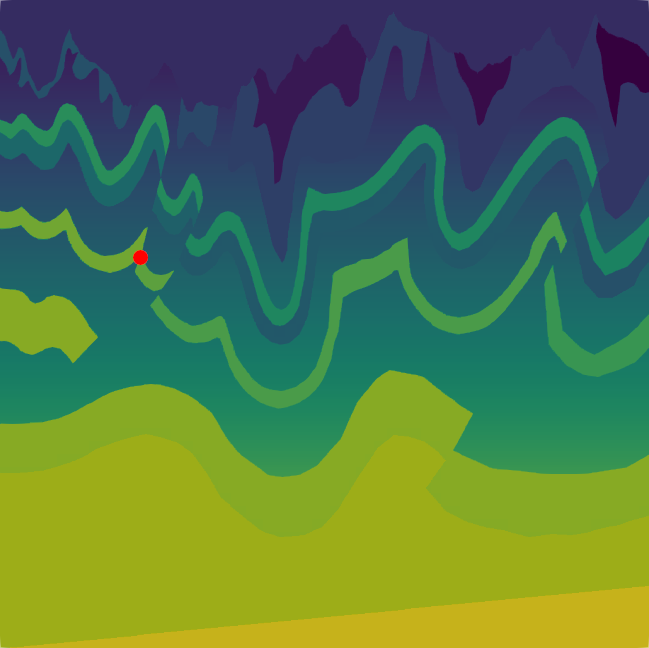}\\[0.5em]
\begin{scaletikzpicturetowidth}{0.75\textwidth}
\begin{tikzpicture}[scale=\tikzscale,font=\large]
\pgfmathsetlengthmacro\MajorTickLength{
  \pgfkeysvalueof{/pgfplots/major tick length} * 0.5
}
\begin{axis}[
title={\Large $\mu$},
xmin=0, xmax=1,
ymin=0, ymax=0.02,
axis on top,
scaled x ticks=false,
scaled y ticks=false,
ytick=\empty,
yticklabels=\empty,
yticklabel pos=right,
x tick label style={
  /pgf/number format/.cd,
                fixed,
        precision=1,
  /tikz/.cd  
},
extra x tick style={
    font=\large,
    tick style=transparent,     yticklabel pos=left,
    x tick label style={
        /pgf/number format/.cd,
            std,
            precision=3,
      /tikz/.cd
    }
},
width=7cm,
height=1.82cm,
major tick length=\MajorTickLength,
max space between ticks=1000pt,
try min ticks=4,
]
\addplot graphics [
includegraphics cmd=\pgfimage,
xmin=\pgfkeysvalueof{/pgfplots/xmin}, 
xmax=\pgfkeysvalueof{/pgfplots/xmax}, 
ymin=\pgfkeysvalueof{/pgfplots/ymin}, 
ymax=\pgfkeysvalueof{/pgfplots/ymax}
] {./figures/colormap_viridis_rotated.png};
\end{axis}
\end{tikzpicture}
\end{scaletikzpicturetowidth}
\end{center}
\end{minipage}
\begin{minipage}{0.32\textwidth}
\begin{center}
\includegraphics[width=\textwidth]{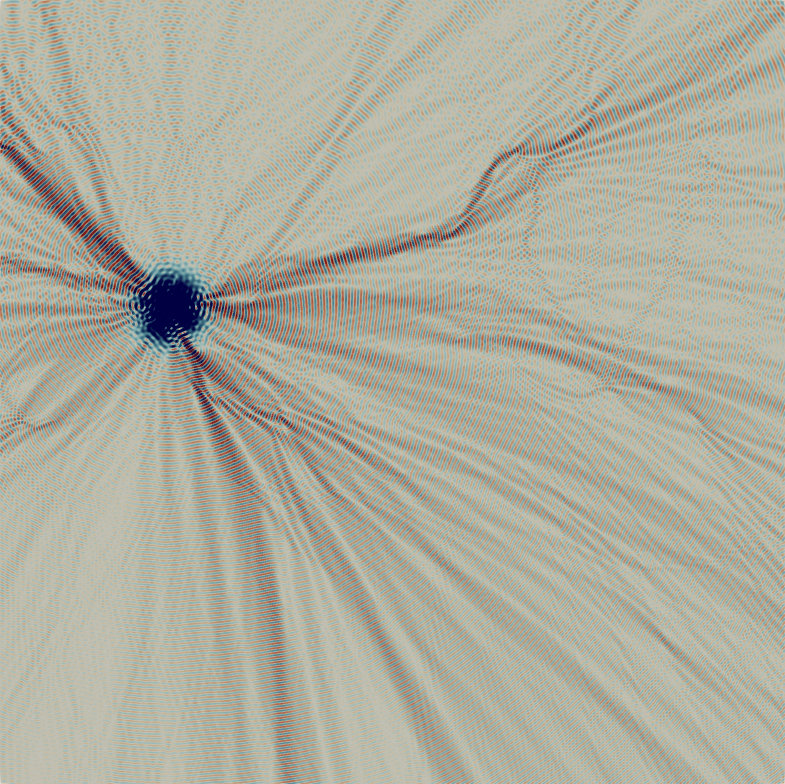}\\[0.5em]
\begin{scaletikzpicturetowidth}{0.75\textwidth}
\begin{tikzpicture}[scale=\tikzscale,font=\large]
\pgfmathsetlengthmacro\MajorTickLength{
  \pgfkeysvalueof{/pgfplots/major tick length} * 0.5
}
\begin{axis}[
title={\Large $\mathrm{Re}(u)$},
xmin=-8e-7, xmax=8e-7,
ymin=0, ymax=0.02,
axis on top,
scaled x ticks=false,
scaled y ticks=false,
ytick=\empty,
yticklabels=\empty,
yticklabel pos=right,
xtick = {0},
xticklabels = {0},
extra x ticks={
   \pgfkeysvalueof{/pgfplots/xmin},
   \pgfkeysvalueof{/pgfplots/xmax}
},
extra x tick style={
    font=\large,
    tick style=transparent,     yticklabel pos=left,
    x tick label style={
        /pgf/number format/.cd,
            std,
      /tikz/.cd
    }
},
width=7cm,
height=1.82cm,
major tick length=\MajorTickLength,
max space between ticks=1000pt,
try min ticks=2,
]
\addplot graphics [
includegraphics cmd=\pgfimage,
xmin=\pgfkeysvalueof{/pgfplots/xmin}, 
xmax=\pgfkeysvalueof{/pgfplots/xmax}, 
ymin=\pgfkeysvalueof{/pgfplots/ymin}, 
ymax=\pgfkeysvalueof{/pgfplots/ymax}
] {./figures/colormap_cool_to_warm_extended_rotated.png};
\end{axis}
\end{tikzpicture}
\end{scaletikzpicturetowidth}
\end{center}
\end{minipage}
\begin{minipage}{0.32\textwidth}
\begin{center}
\includegraphics[width=\textwidth]{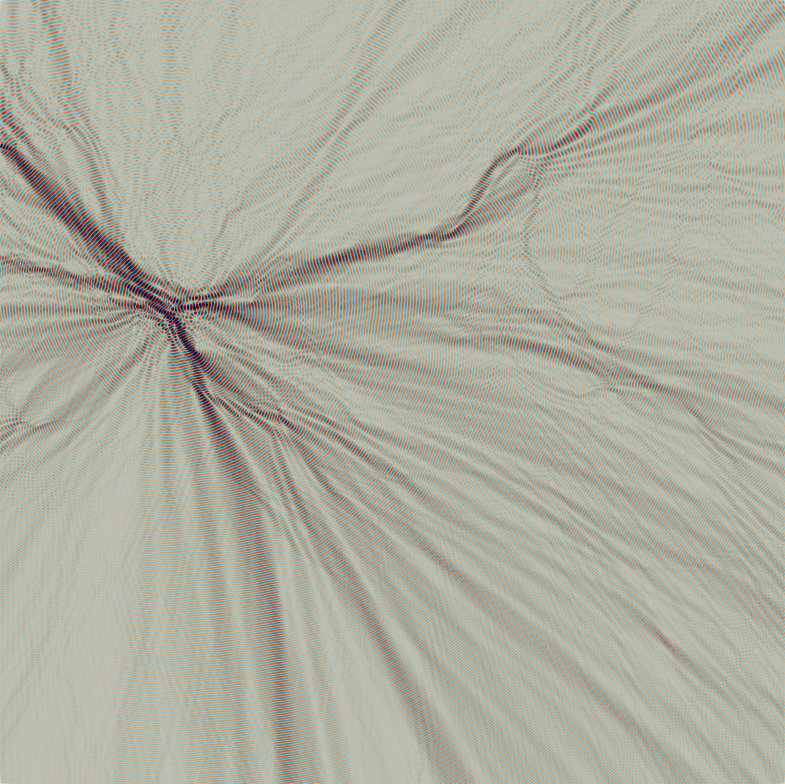}\\[0.5em]
\begin{scaletikzpicturetowidth}{0.75\textwidth}
\begin{tikzpicture}[scale=\tikzscale,font=\large]
\pgfmathsetlengthmacro\MajorTickLength{
  \pgfkeysvalueof{/pgfplots/major tick length} * 0.5
}
\begin{axis}[
title={\Large $\mathrm{Im}(u)$},
xmin=-8e-7, xmax=8e-7,
ymin=0, ymax=0.02,
axis on top,
scaled x ticks=false,
scaled y ticks=false,
ytick=\empty,
yticklabels=\empty,
yticklabel pos=right,
xtick = {0},
xticklabels = {0},
extra x ticks={
   \pgfkeysvalueof{/pgfplots/xmin},
   \pgfkeysvalueof{/pgfplots/xmax}
},
extra x tick style={
    font=\large,
    tick style=transparent,     yticklabel pos=left,
    x tick label style={
        /pgf/number format/.cd,
            std,
      /tikz/.cd
    }
},
width=7cm,
height=1.82cm,
]
\addplot graphics [
includegraphics cmd=\pgfimage,
xmin=\pgfkeysvalueof{/pgfplots/xmin}, 
xmax=\pgfkeysvalueof{/pgfplots/xmax}, 
ymin=\pgfkeysvalueof{/pgfplots/ymin}, 
ymax=\pgfkeysvalueof{/pgfplots/ymax}
] {./figures/colormap_cool_to_warm_extended_rotated.png};
\end{axis}
\end{tikzpicture}
\end{scaletikzpicturetowidth}
\end{center}
\end{minipage}
\caption{\label{fig:bpmigration2dfigures} Migration from topography velocity profile with source location (left) and real and imaginary part of the solution in the case $p=2$ (center and right).}
\end{figure}
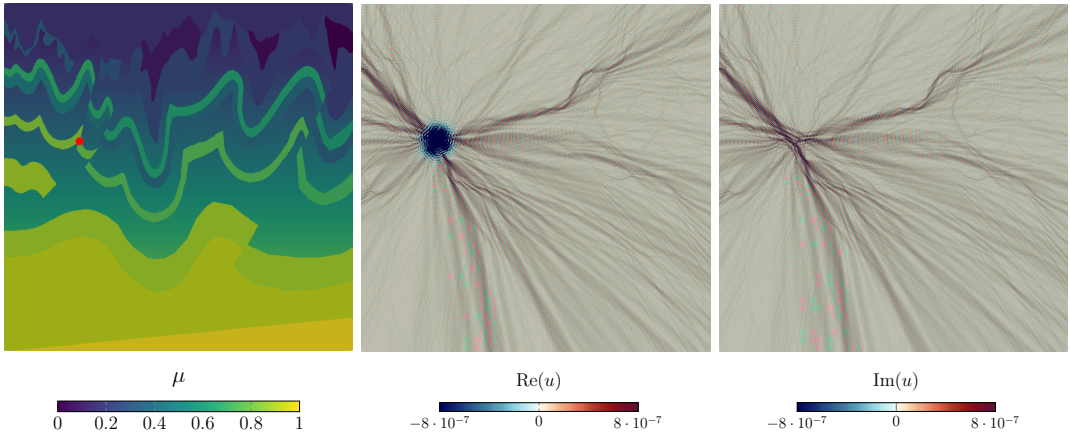
In a third scenario, we consider the velocity profile stemming from the migration from topography model representing a cross section through the foothills of the Canadian rockies; courtesy of Amoco and BP \cite{Gray1995Migration}.
The corresponding scaled velocity profile is illustrated in the left of \cref{fig:bpmigration2dfigures}.
Here, the maximum value of the source term is located at $\bms = (0.2159, 0.6054)^\T$.
We collect the required number of FGMRES iterations and the respective compute times for $p \in \{1,2,3\}$ and the maximum wavenumbers $k_{\mathrm{max}} \in \{500, 1100, 1900\}$ in \cref{tab:bpmigration2d}.
Additionally, in the center and right of \cref{fig:bpmigration2dfigures}, we present the real and imaginary parts of the solution in the case of $p=2$.
We observe that using the optimal shift exponent results in the smallest number of iterations and the shortest compute time throughout all the considered examples.
The choice $k^{\frac{3}{2}}$ yields similar small iterations when compared to using no shift or a shift of $k$, but using the trained shift yields the fastest solution without requiring a manual choice of the shift.

\begin{table}[H]
\centering
\footnotesize
\caption{\label{tab:bpmigration2d}Parameter values, iteration numbers, and compute times for the 2D migration from topography example.}
\begin{tabular}{ll}
\begin{tabular}[t]{l|l}
Param. & Value\\
\hline
$p$ & $1$\\
$h$ & $2^{-10}$\\
$k_{\mathrm{max}}$ & $500$\\
$k_{\mathrm{max}} \cdot h$ & $0.4883$\\
DoFs & $2\,101\,250$
\end{tabular}
&
\begin{tabular}[t]{c|r|r|r}
$\varepsilon$ & Iter & Time [m:s] & Speed-up \\
\hline
$0$               & 421  & 3:38.80 &  \\
$k$               & 345  & 2:18.51 & 57.97\% \\
$k^{\frac{3}{2}}$ & 118  & 0:40.02 & 446.73\% \\
$k^{2}$           & >500 & --      & -- \\
$k^\shiftexp$     & 107  & 0:35.98 & 508.12\%
\end{tabular}\vspace*{0.5em}\\
\begin{tabular}[t]{l|l}
Param. & Value\\
\hline
$p$ & $2$\\
$h$ & $2^{-10}$\\
$k_{\mathrm{max}}$ & $1100$\\
$k_{\mathrm{max}} \cdot h$ & $1.07422$\\
DoFs & $8\,396\,802$
\end{tabular}
&
\begin{tabular}[t]{c|r|r|r}
$\varepsilon$ & Iter & Time [m:s] & Speed-up \\
\hline
$0$               &  147 & 5:09.24 &  \\
$k$               &  134 & 4:05.64 & 25.89\% \\
$k^{\frac{3}{2}}$ &   91 & 2:36.09 & 98.12\% \\
$k^{2}$           & >500 & --      & -- \\
$k^\shiftexp$     &   73 & 2:18.60 & 123.12\%
\end{tabular}\vspace*{0.5em}\\

\begin{tabular}[t]{l|l}
Param. & Value\\
\hline
$p$ & $3$\\
$h$ & $2^{-10}$\\
$k_{\mathrm{max}}$ & $1900$\\
$k_{\mathrm{max}} \cdot h$ & $1.85547$\\
DoFs & $18\,886\,658$
\end{tabular}
&
\begin{tabular}[t]{c|r|r|r}
$\varepsilon$ & Iter & Time [m:s] & Speed-up \\
\hline
$0$               &  402 & 31:39.68 &  \\
$k$               &  345 & 26:23.38 & 19.98\% \\
$k^{\frac{3}{2}}$ &  118 &  7:48.05 & 305.87\% \\
$k^{2}$           & >500 & --       & -- \\
$k^\shiftexp$     &  108 &  7:04.33 & 347.69\%
\end{tabular}
\end{tabular}
\end{table}

\subsection{BP statics benchmark model}
\begin{figure}[H]
\centering
\begin{minipage}{0.32\textwidth}
\begin{center}
\includegraphics[width=\textwidth]{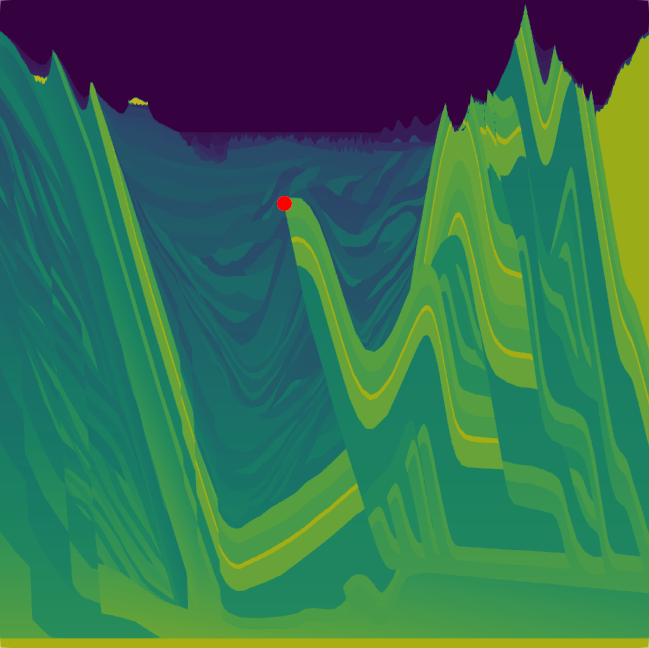}\\[0.5em]
\begin{scaletikzpicturetowidth}{0.75\textwidth}
\begin{tikzpicture}[scale=\tikzscale,font=\large]
\pgfmathsetlengthmacro\MajorTickLength{
  \pgfkeysvalueof{/pgfplots/major tick length} * 0.5
}
\begin{axis}[
title={\Large $\mu$},
xmin=0, xmax=1,
ymin=0, ymax=0.02,
axis on top,
scaled x ticks=false,
scaled y ticks=false,
ytick=\empty,
yticklabels=\empty,
yticklabel pos=right,
x tick label style={
  /pgf/number format/.cd,
                fixed,
        precision=1,
  /tikz/.cd  
},
extra x tick style={
    font=\large,
    tick style=transparent,     yticklabel pos=left,
    x tick label style={
        /pgf/number format/.cd,
            std,
            precision=3,
      /tikz/.cd
    }
},
width=7cm,
height=1.82cm,
major tick length=\MajorTickLength,
max space between ticks=1000pt,
try min ticks=4,
]
\addplot graphics [
includegraphics cmd=\pgfimage,
xmin=\pgfkeysvalueof{/pgfplots/xmin}, 
xmax=\pgfkeysvalueof{/pgfplots/xmax}, 
ymin=\pgfkeysvalueof{/pgfplots/ymin}, 
ymax=\pgfkeysvalueof{/pgfplots/ymax}
] {./figures/colormap_viridis_rotated.png};
\end{axis}
\end{tikzpicture}
\end{scaletikzpicturetowidth}
\end{center}
\end{minipage}
\begin{minipage}{0.32\textwidth}
\begin{center}
\includegraphics[width=\textwidth]{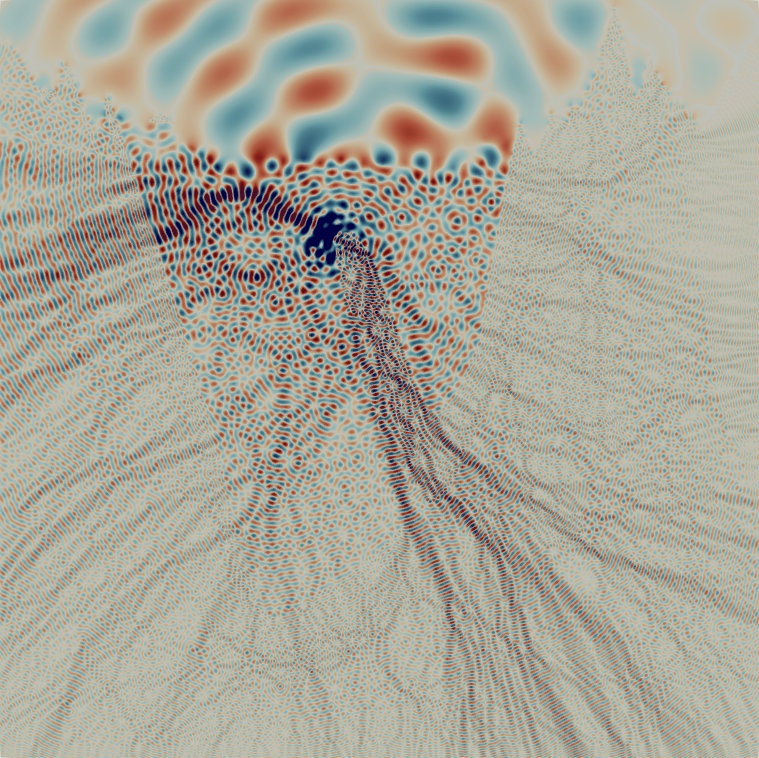}\\[0.5em]
\begin{scaletikzpicturetowidth}{0.75\textwidth}
\begin{tikzpicture}[scale=\tikzscale,font=\large]
\pgfmathsetlengthmacro\MajorTickLength{
  \pgfkeysvalueof{/pgfplots/major tick length} * 0.5
}
\begin{axis}[
title={\Large $\mathrm{Re}(u)$},
xmin=-7e-6, xmax=7e-6,
ymin=0, ymax=0.02,
axis on top,
scaled x ticks=false,
scaled y ticks=false,
ytick=\empty,
yticklabels=\empty,
yticklabel pos=right,
xtick = {0},
xticklabels = {0},
extra x ticks={
   \pgfkeysvalueof{/pgfplots/xmin},
   \pgfkeysvalueof{/pgfplots/xmax}
},
extra x tick style={
    font=\large,
    tick style=transparent,     yticklabel pos=left,
    x tick label style={
        /pgf/number format/.cd,
            std,
      /tikz/.cd
    }
},
width=7cm,
height=1.82cm,
major tick length=\MajorTickLength,
max space between ticks=1000pt,
try min ticks=2,
]
\addplot graphics [
includegraphics cmd=\pgfimage,
xmin=\pgfkeysvalueof{/pgfplots/xmin}, 
xmax=\pgfkeysvalueof{/pgfplots/xmax}, 
ymin=\pgfkeysvalueof{/pgfplots/ymin}, 
ymax=\pgfkeysvalueof{/pgfplots/ymax}
] {./figures/colormap_cool_to_warm_extended_rotated.png};
\end{axis}
\end{tikzpicture}
\end{scaletikzpicturetowidth}
\end{center}
\end{minipage}
\begin{minipage}{0.32\textwidth}
\begin{center}
\includegraphics[width=\textwidth]{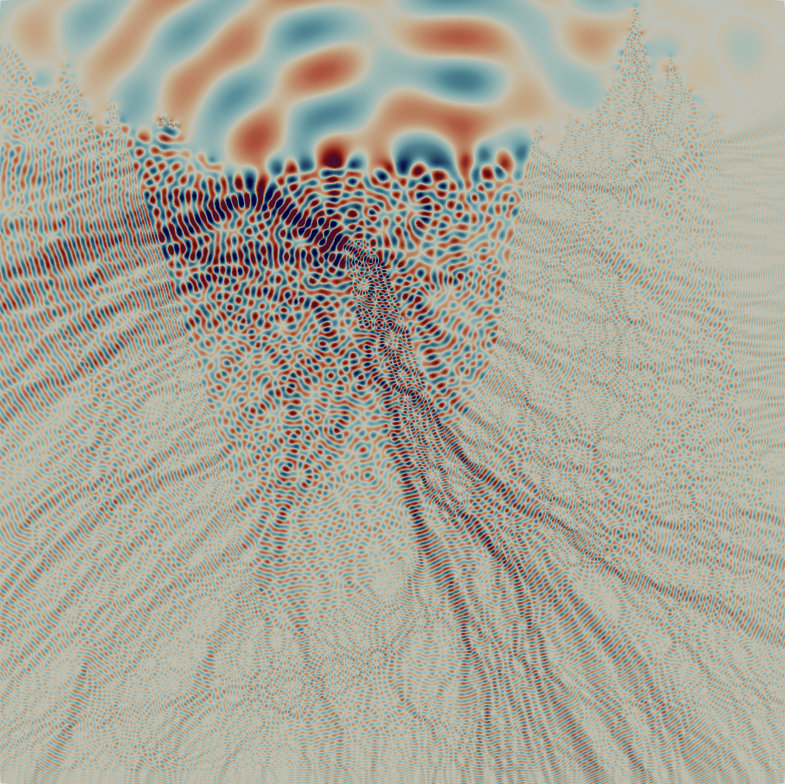}\\[0.5em]
\begin{scaletikzpicturetowidth}{0.75\textwidth}
\begin{tikzpicture}[scale=\tikzscale,font=\large]
\pgfmathsetlengthmacro\MajorTickLength{
  \pgfkeysvalueof{/pgfplots/major tick length} * 0.5
}
\begin{axis}[
title={\Large $\mathrm{Im}(u)$},
xmin=-7e-6, xmax=7e-6,
ymin=0, ymax=0.02,
axis on top,
scaled x ticks=false,
scaled y ticks=false,
ytick=\empty,
yticklabels=\empty,
yticklabel pos=right,
xtick = {0},
xticklabels = {0},
extra x ticks={
   \pgfkeysvalueof{/pgfplots/xmin},
   \pgfkeysvalueof{/pgfplots/xmax}
},
extra x tick style={
    font=\large,
    tick style=transparent,     yticklabel pos=left,
    x tick label style={
        /pgf/number format/.cd,
            std,
      /tikz/.cd
    }
},
width=7cm,
height=1.82cm,
]
\addplot graphics [
includegraphics cmd=\pgfimage,
xmin=\pgfkeysvalueof{/pgfplots/xmin}, 
xmax=\pgfkeysvalueof{/pgfplots/xmax}, 
ymin=\pgfkeysvalueof{/pgfplots/ymin}, 
ymax=\pgfkeysvalueof{/pgfplots/ymax}
] {./figures/colormap_cool_to_warm_extended_rotated.png};
\end{axis}
\end{tikzpicture}
\end{scaletikzpicturetowidth}
\end{center}
\end{minipage}
\caption{\label{fig:bpstatics2dfigures} BP statics benchmark model velocity profile with source location (left) and real and imaginary part of the solution in the case $p=2$ (center and right).}
\end{figure}
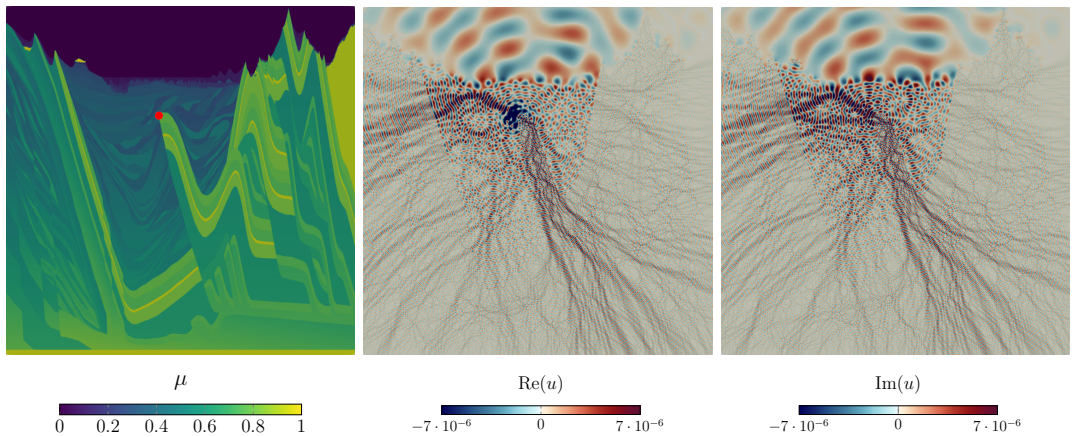
In a fourth scenario, we consider the velocity profile stemming from the synthetic BP statics benchmark model created by Mike O'Brien and Carl Regone, provided by courtesy of Amoco and BP \cite{AmocoStatics}.
The corresponding scaled velocity profile is illustrated in the left of \cref{fig:bpstatics2dfigures}.
Here, the maximum value of the source term is located at $\bms = (0.4368, 0.6852)^\T$.
We collect the required number of FGMRES iterations and the respective compute times for $p \in \{1,2,3\}$ and the maximum wavenumbers $k_{\mathrm{max}} \in \{450, 1100, 1900\}$ in \cref{tab:bpstatics2d}.
Additionally, in the center and right of \cref{fig:bpstatics2dfigures}, we present the real and imaginary parts of the solution in the case of $p=2$.
We observe that using the optimal shift exponent results in the smallest number of iterations and shortest compute time throughout all the considered examples.
Considerably, for $p=2$ and $p=3$ the choice of $k^{\frac{2}{3}}$ results in a larger number of required iterations and a longer compute time than for the case without a shift.

\begin{table}[H]
\centering
\footnotesize
\caption{\label{tab:bpstatics2d}Parameter values, iteration numbers, and compute times for the 2D BP statics benchmark model example.}
\begin{tabular}{ll}
\begin{tabular}[t]{l|l}
Param. & Value\\
\hline
$p$ & $1$\\
$h$ & $2^{-10}$\\
$k_{\mathrm{max}}$ & 450\\
$k_{\mathrm{max}} \cdot h$ & 0.439453\\
DoFs & $2\,101\,250$
\end{tabular}
&
\begin{tabular}[t]{c|r|r|r}
$\varepsilon$ & Iter & Time [m:s] & Speed-up \\
\hline
$0$               & 192 & 1:35.43 & -- \\
$k$               & 154 & 0:59.33 & 60.85\% \\
$k^{\frac{3}{2}}$ & 125 & 0:42.00 & 127.21\% \\
$k^{2}$           &>500 & --      & -- \\
$k^\shiftexp$     & 109 & 0:39.22 & 143.32\%
\end{tabular}\vspace{0.5em}\\

\begin{tabular}[t]{l|l}
Param. & Value\\
\hline
$p$ & $2$\\
$h$ & $2^{-10}$\\
$k_{\mathrm{max}}$ & $1100$\\
$k_{\mathrm{max}} \cdot h$ & $1.07422$\\
DoFs & $8\,396\,802$
\end{tabular}
&
\begin{tabular}[t]{c|r|r|r}
$\varepsilon$ & Iter & Time [m:s] & Speed-up \\
\hline
$0$               &  113 & 4:01.63 & -- \\
$k$               &   94 & 2:53.50 & \% \\
$k^{\frac{3}{2}}$ &  148 & 4:25.31 & -8.93\% \\
$k^{2}$           & >500 & --      & -- \\
$k^\shiftexp$     &   79 & 2:27.00 & 64.37\%
\end{tabular}\vspace*{0.5em}\\

\begin{tabular}[t]{l|l}
Param. & Value\\
\hline
$p$ & $3$\\
$h$ & $2^{-10}$\\
$k_{\mathrm{max}}$ & $1900$\\
$k_{\mathrm{max}} \cdot h$ & $1.85547$\\
DoFs & $18\,886\,658$
\end{tabular}
&
\begin{tabular}[t]{c|r|r|r}
$\varepsilon$ & Iter & Time [m:s] & Speed-up \\
\hline
$0$               &  146 &  9:58.49 & -- \\
$k$               &  129 &  8:28.25 & 17.76\% \\
$k^{\frac{3}{2}}$ &  185 & 12:23.17 & -19.47\% \\
$k^{2}$           & >500 & --      & -- \\
$k^\shiftexp$     &  107 &  7:01.24 & 42.08\%
\end{tabular}
\end{tabular}
\end{table}

\subsection{Marmousi 3D model}
\begin{figure}[H]
\centering
\begin{minipage}{0.32\textwidth}
\begin{center}
\includegraphics[width=\textwidth]{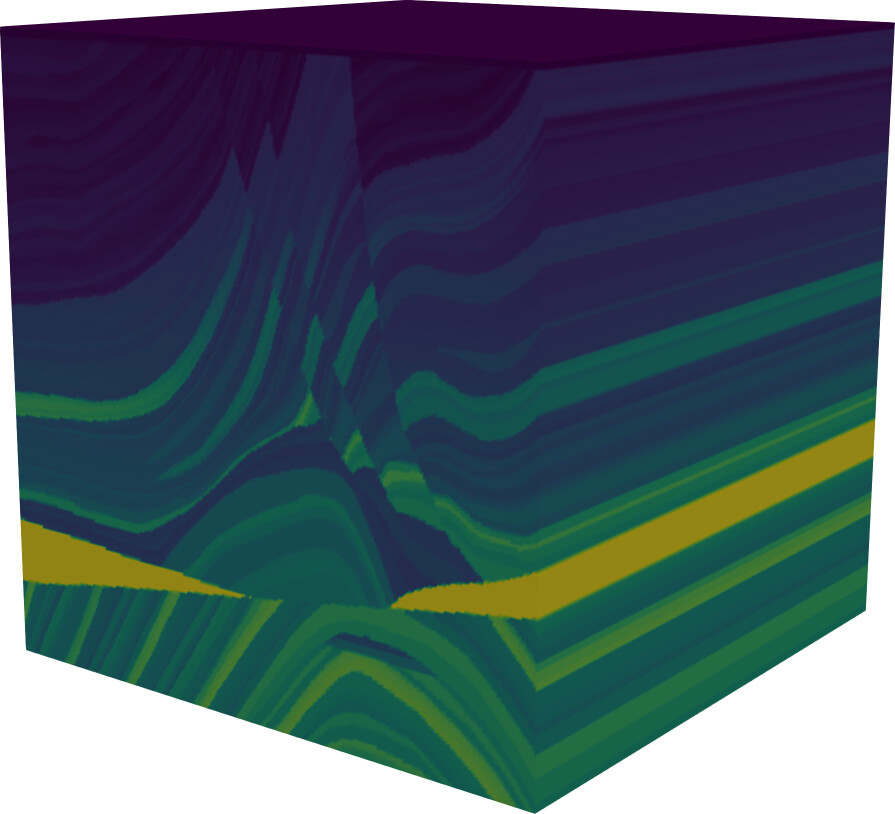}\\[0.5em]
\begin{scaletikzpicturetowidth}{0.75\textwidth}
\begin{tikzpicture}[scale=\tikzscale,font=\large]
\pgfmathsetlengthmacro\MajorTickLength{
  \pgfkeysvalueof{/pgfplots/major tick length} * 0.5
}
\begin{axis}[
title={\Large $\mu$},
xmin=0, xmax=1,
ymin=0, ymax=0.02,
axis on top,
scaled x ticks=false,
scaled y ticks=false,
ytick=\empty,
yticklabels=\empty,
yticklabel pos=right,
x tick label style={
  /pgf/number format/.cd,
                fixed,
        precision=1,
  /tikz/.cd  
},
extra x tick style={
    font=\large,
    tick style=transparent,     yticklabel pos=left,
    x tick label style={
        /pgf/number format/.cd,
            std,
            precision=3,
      /tikz/.cd
    }
},
width=7cm,
height=1.82cm,
major tick length=\MajorTickLength,
max space between ticks=1000pt,
try min ticks=4,
]
\addplot graphics [
includegraphics cmd=\pgfimage,
xmin=\pgfkeysvalueof{/pgfplots/xmin}, 
xmax=\pgfkeysvalueof{/pgfplots/xmax}, 
ymin=\pgfkeysvalueof{/pgfplots/ymin}, 
ymax=\pgfkeysvalueof{/pgfplots/ymax}
] {./figures/colormap_viridis_rotated.png};
\end{axis}
\end{tikzpicture}
\end{scaletikzpicturetowidth}
\end{center}
\end{minipage}
\begin{minipage}{0.32\textwidth}
\begin{center}
\includegraphics[width=\textwidth]{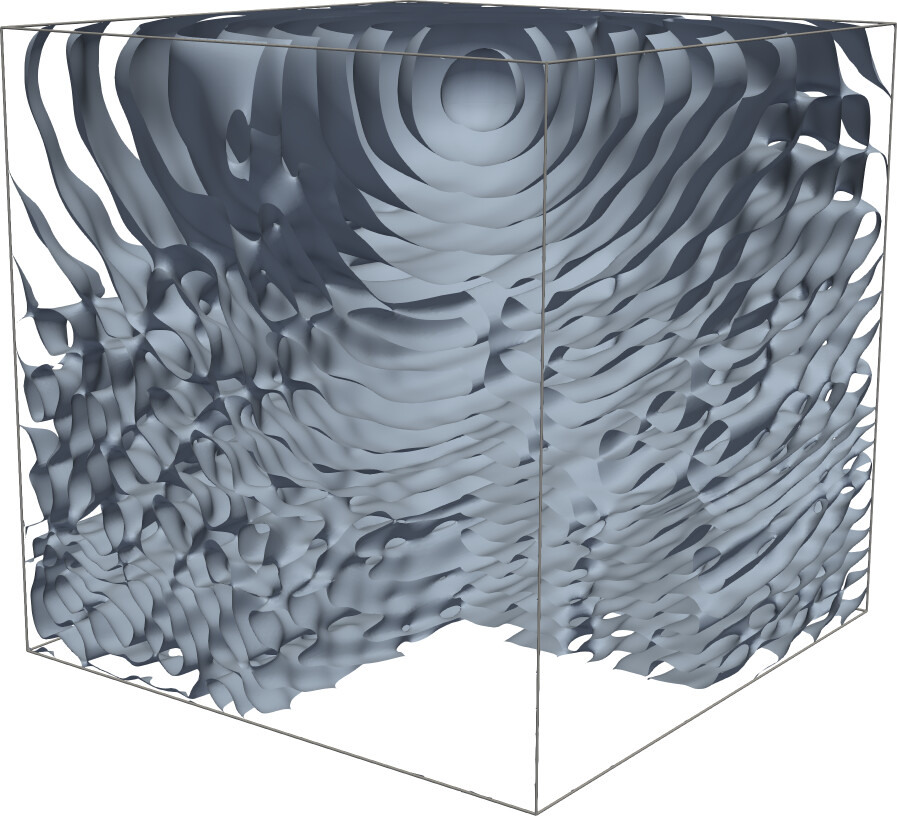}\\[0.5em]
\phantom{\begin{scaletikzpicturetowidth}{0.75\textwidth}
\begin{tikzpicture}[scale=\tikzscale,font=\large]
\pgfmathsetlengthmacro\MajorTickLength{
  \pgfkeysvalueof{/pgfplots/major tick length} * 0.5
}
\begin{axis}[
title={\Large $\mu$},
xmin=0, xmax=1,
ymin=0, ymax=0.02,
axis on top,
scaled x ticks=false,
scaled y ticks=false,
ytick=\empty,
yticklabels=\empty,
yticklabel pos=right,
x tick label style={
  /pgf/number format/.cd,
                fixed,
        precision=1,
  /tikz/.cd  
},
extra x tick style={
    font=\large,
    tick style=transparent,     yticklabel pos=left,
    x tick label style={
        /pgf/number format/.cd,
            std,
            precision=3,
      /tikz/.cd
    }
},
width=7cm,
height=1.82cm,
major tick length=\MajorTickLength,
max space between ticks=1000pt,
try min ticks=4,
]
\addplot graphics [
includegraphics cmd=\pgfimage,
xmin=\pgfkeysvalueof{/pgfplots/xmin}, 
xmax=\pgfkeysvalueof{/pgfplots/xmax}, 
ymin=\pgfkeysvalueof{/pgfplots/ymin}, 
ymax=\pgfkeysvalueof{/pgfplots/ymax}
] {./figures/colormap_viridis_rotated.png};
\end{axis}
\end{tikzpicture}
\end{scaletikzpicturetowidth}}
\end{center}
\end{minipage}
\begin{minipage}{0.32\textwidth}
\begin{center}
\includegraphics[width=\textwidth]{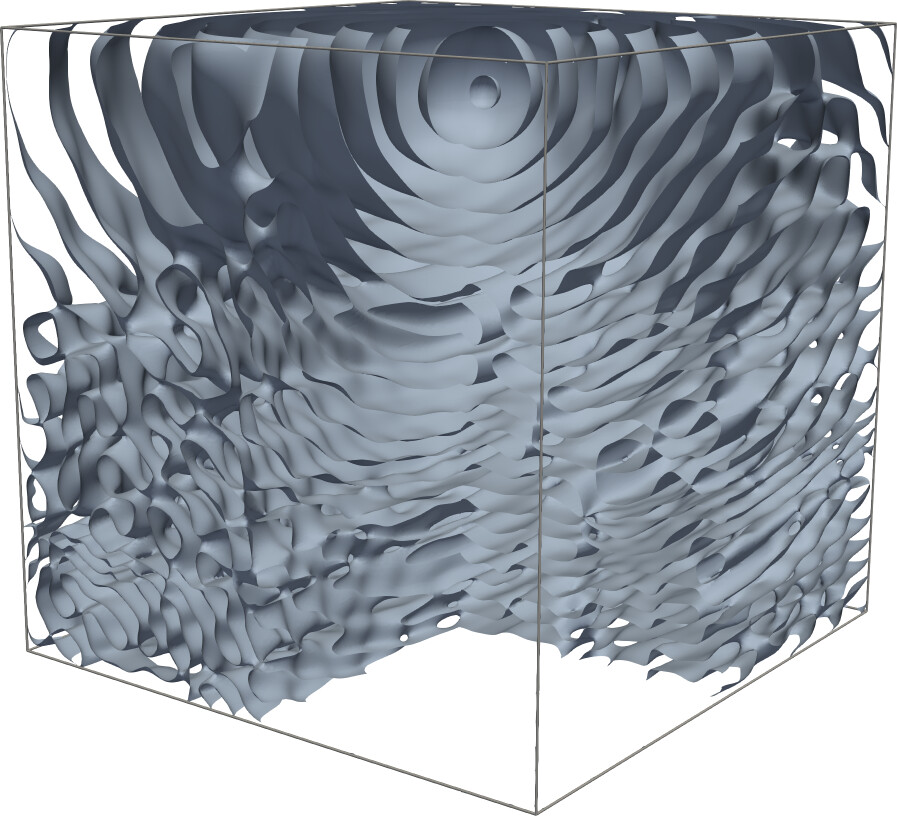}\\[0.5em]
\phantom{\begin{scaletikzpicturetowidth}{0.75\textwidth}
\begin{tikzpicture}[scale=\tikzscale,font=\large]
\pgfmathsetlengthmacro\MajorTickLength{
  \pgfkeysvalueof{/pgfplots/major tick length} * 0.5
}
\begin{axis}[
title={\Large $\mu$},
xmin=0, xmax=1,
ymin=0, ymax=0.02,
axis on top,
scaled x ticks=false,
scaled y ticks=false,
ytick=\empty,
yticklabels=\empty,
yticklabel pos=right,
x tick label style={
  /pgf/number format/.cd,
                fixed,
        precision=1,
  /tikz/.cd  
},
extra x tick style={
    font=\large,
    tick style=transparent,     yticklabel pos=left,
    x tick label style={
        /pgf/number format/.cd,
            std,
            precision=3,
      /tikz/.cd
    }
},
width=7cm,
height=1.82cm,
major tick length=\MajorTickLength,
max space between ticks=1000pt,
try min ticks=4,
]
\addplot graphics [
includegraphics cmd=\pgfimage,
xmin=\pgfkeysvalueof{/pgfplots/xmin}, 
xmax=\pgfkeysvalueof{/pgfplots/xmax}, 
ymin=\pgfkeysvalueof{/pgfplots/ymin}, 
ymax=\pgfkeysvalueof{/pgfplots/ymax}
] {./figures/colormap_viridis_rotated.png};
\end{axis}
\end{tikzpicture}
\end{scaletikzpicturetowidth}}
\end{center}
\end{minipage}
\caption{\label{fig:marmousi3dfigures} Marmousi 3D model velocity profile (left) and isosurfaces at value $0$ of the real and imaginary part of the solution in the case $p=1$ (center and right).}
\end{figure}
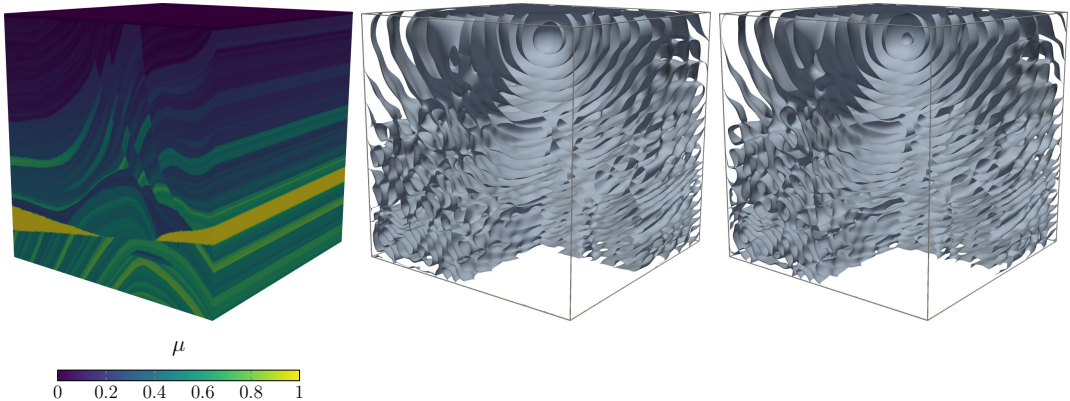

In the last scenario, we consider again the velocity profile stemming from the synthetic Marmousi model devised by the Institut Fran\c{c}ais du Petrole \cite{Versteeg1994Marmousi} extruded along a third dimension.
The corresponding scaled velocity profile is illustrated in the left of \cref{fig:marmousi3dfigures}.
Here, the maximum value of the source term is located at $\bms = (0.5421, 0.8946, 0.5)^\T$.
We collect the required number of FGMRES iterations and the respective compute times for $p \in \{1,2\}$ and the maximum wavenumbers $k_{\mathrm{max}} = 150$ in \cref{tab:marmousi3d}.
Additionally, in the center and right of \cref{fig:marmousi3dfigures}, we present the real and imaginary parts of the solution for $p=1$.
The timings were obtained on the SuperMUC-NG cluster described above by using 384 cores across 16 compute nodes,i.e.,~24 cores per compute node.
We observe that for $p=1$, the near optimal shift yields the smallest number of outer FGMRES iterations, but like in the 2D Marmousi case, using a shift of $k^{\frac{3}{2}}$ is faster even if six more iterations are performed.
For $p=2$, the number of iterations using the near optimal shift was larger by one compared to the smallest number of iterations obtained by using no shift or a shift by $k$, but the compute time was still smaller.
These discrepancies in the run-time and number of iterations may be explained by the time deviations of the parallel LU decomposition performed in the first application of the preconditioner.

\begin{table}[H]
\centering
\footnotesize
\caption{\label{tab:marmousi3d}Parameter values, iteration numbers, and compute times for the 3D Marmousi model example with $p=1$.}
\begin{tabular}{ll}
\begin{tabular}[t]{l|l}
Param. & Value\\
\hline
$p$ & $1$\\
$h$ & $2^{-8}$\\
$k_{\mathrm{max}}$ & 150\\
$k_{\mathrm{max}} \cdot h$ & 0.585938\\
DoFs & $33\,949\,186$
\end{tabular}
&
\begin{tabular}[t]{c|r|r|r}
$\varepsilon$ & Iter & Time [m:s] & Speed-up \\
\hline
$0$               & 51 & 12:47.05 & -- \\
$k$               & 47 & 12:03.38 & 6.04\% \\
$k^{\frac{3}{2}}$ & 45 & 10:40.04 & 19.84\% \\
$k^{2}$           &239 & 25:03.34 & -48.98\% \\
$k^\shiftexp$     & 39 & 11:17.22 & 13.26\%
\end{tabular}\vspace*{0.5em}\\
\begin{tabular}[t]{l|l}
Param. & Value\\
\hline
$p$ & $2$\\
$h$ & $2^{-7}$\\
$k_{\mathrm{max}}$ & 150\\
$k_{\mathrm{max}} \cdot h$ & 1.17188\\
DoFs & $33\,949\,186$
\end{tabular}
&
\begin{tabular}[t]{c|r|r|r}
$\varepsilon$ & Iter & Time [m:s] & Speed-up \\
\hline
$0$               & 16 & 10:35.21 & -- \\
$k$               & 16 & 10:01.42 & 5.62\% \\
$k^{\frac{3}{2}}$ & 35 & 10:56.22 & -3.20\% \\
$k^{2}$           &241 & 24:13.33 & -56.29\% \\
$k^\shiftexp$     & 17 &  9:59.29 & 5.99\%
\end{tabular}
\end{tabular}
\end{table}
\section{Conclusion}
\label{sec:Conclusion}
In this work, we have presented a preconditioner for the Helmholtz equation obtained from a data driven approach.
The preconditioner uses near optimal complex shifts in the shifted Laplacian problem which is used as a preconditioner of the Helmholtz equation by applying a twogrid V-cycle to the discrete problem.
The near optimal shifts were obtained by generating training data for different mesh sizes $h$, wavenumbers $k$, and discretization orders $p$, and subsequently performing a nonlinear regression to construct a near optimal shift map.
Using such an approximated optimal shift map allows users to obtain near optimal shifts automatically without having to tune the required complex shifts manually.
In order to solidify this approach, we have performed theoretical considerations based on a local Fourier analysis which justify this data driven approach and we have related the theoretical results to experimental data.
Additionally, the twogrid method has been implemented in a semi matrix-free fashion which saves on memory storage and traffic which usually required for matrices corresponding to the finer grids.
Furthermore, we have used these near optimal shifts on a set of numerical benchmarks with heterogeneous wavenumbers in 2D and 3D.
It could be observed that using these near optimal shifts yielded the smallest FGMRES iteration numbers throughout almost all the examples with speed ups up to $582$\%.

In the data generation and the numerical experiments, we had restricted ourselves to a single $V(3,3)$ twogrid cycle and a damped Jacobi smoother with damping factor $\omega = \frac{2}{3}$.
Moreover, the complex shift had been always in the form $\zI k^\shiftexp$.
Possible further work could take into account different parameters for the multigrid solver and wavenumber coefficients of the form $\beta_1 k^{\shiftexp_1} + \zI \beta_2 k^{\shiftexp_2}$ with $\beta_1, \beta_2, \shiftexp_1, \shiftexp_2 \in \R$.

\section*{Acknowledgments}
This work was partly supported by the German Research Foundation by grant WO671/11-1.
The authors gratefully acknowledge the Gauss Centre for Supercomputing e.V. (\url{www.gauss-centre.eu}) for funding this project by providing computing time on the GCS Supercomputer SuperMUC at Leibniz Supercomputing Centre (\url{www.lrz.de}).
\FloatBarrier
\phantomsection\bibliographystyle{abbrv}
\bibliography{main}

\end{document}